\documentclass[final,authoryear,3p,times,twocolumn]{elsarticle}
\usepackage[utf8]{inputenc}

\usepackage{graphicx}
\usepackage{amssymb}
\pdfoutput=1

\usepackage[colorlinks = true,
            citecolor = blue]{hyperref}
\usepackage{gensymb}

\journal{Icarus}

\begin{document}

\begin{frontmatter}


\title{Longitudinal Variations in the Stratosphere of Uranus from the Spitzer Infrared Spectrometer}



\author[label1]{Naomi Rowe-Gurney}
\author[label1]{Leigh N. Fletcher}
\author[label2]{Glenn S. Orton}
\author[label1]{Michael T. Roman}
\author[label3]{Amy Mainzer}
\author[label4]{Julianne I. Moses}
\author[label5]{Imke de Pater}
\author[label6]{Patrick G. J. Irwin}

\address[label1]{School of Physics and Astronomy, University of Leicester, Leicester, United Kingdom} 
\address[label2]{Jet Propulsion Laboratory, California Institute of Technology, Pasadena, California, United States\\}
\address[label3]{Lunar and Planetary Laboratory, The University of Arizona, Tucson, Arizona, United States}
\address[label4]{Space Science Institute, Seabrook, Texas, United States}
\address[label5]{University of California at Berkeley, Berkeley, California, United States}
\address[label6]{Atmospheric, Oceanic, and Planetary Physics, University of Oxford, Oxford, United Kingdom}

\begin{abstract}
NASA's Spitzer Infrared Spectrometer (IRS) acquired mid-infrared (5-37 $\mu$m) disc-averaged spectra of Uranus very near to its equinox in December 2007. A mean spectrum was constructed from observations of multiple central meridian longitudes, spaced equally around the planet, which has provided the opportunity for the most comprehensive globally-averaged characterisation of Uranus' temperature and composition ever obtained \citep{orton2014mid1,orton2014mid2}. In this work we analyse the disc-averaged spectra at four separate central meridian longitudes to reveal significant longitudinal variability in thermal emission occurring in Uranus' stratosphere during the 2007 equinox. We detect a variability of up to 15\% at wavelengths sensitive to stratospheric methane, ethane and acetylene at the $\sim0.1$-mbar level. The tropospheric hydrogen-helium continuum and deuterated methane absorption exhibit a negligible variation (less than 2\%), constraining the phenomenon to the stratosphere. Building on the forward-modelling analysis of the global average study, we present full optimal estimation inversions \cite[using the NEMESIS retrieval algorithm,][]{irwin2008nemesis} of the Uranus-2007 spectra at each longitude to distinguish between thermal and compositional variability. We found that the variations can be explained by a temperature change of less than 3 K in the stratosphere. Near-infrared observations from Keck II NIRC2 in December 2007 \citep{sromovsky2009,depater2011}, and mid-infrared observations from VLT/VISIR in 2009 \citep{roman2020}, help to localise the potential sources to either large scale uplift or stratospheric wave phenomena.
\end{abstract}

%%keywords

\end{frontmatter}

%%\linenumbers

\section{Introduction}
\label{S:intro}

Methane and its photolysis products dominate the composition and temperature structure of Uranus' middle atmosphere \citep{moses2018}. The complex hydrocarbons produced in solar-driven reactions in the stratosphere are the main trace gases present in the stratosphere and upper troposphere. These species are observable at mid-infrared wavelengths sensitive to emission from altitudes between approximately one nanobar and two bars of pressure \citep{orton2014mid1}. 

Uranus' meteorology was perceived to be relatively dormant during the Voyager 2 fly-by in 1986 but has since then increased in activity as Uranus approached its northern spring equinox in 2007, as shown most prominently at near-infrared wavelengths. Episodic bright and dark features were observed in 2011 that were changing and moving over relatively short timescales \citep{sromovsky2012a} and bright, long-lived cloud features have been observed multiple times \citep{depater2011,sromovsky2009,roman2018}. One of the largest and brightest of these features was called the ``Bright Northern Complex'', which attained its peak brightness in 2005 with clouds reaching pressures as low as 240 to 300 mbar \citep{sromovsky2007,roman2018}. In 2014 a similarly bright feature was observed in the near-infrared and estimated to reach to similar heights \citep{depater2015}. These features may be tied to vortex systems that exist in the upper troposphere, such as the prominent dark spot observed in 2006 at depths in the 1-4 bar pressure range \citep{hammel2009}. This feature had bright cloud companions manifesting at lower pressures of around 220 mbars \citep{sromovskyfry2005}, which could be evidence of deep-seated features influencing the structure of the upper troposphere at certain longitudes. Vigorous upwelling of hot gases from the deep atmosphere of giant planets can cause changes in both the temperature and chemistry of the stratosphere, as has been observed at Saturn. A large stratospheric anomaly called the Saturn stratospheric beacon, observed in 2011-2012, was likely due to waves propagating into the stratosphere from a tropospheric storm, interacting with the background zonal flow, breaking and depositing their energy \citep{Fletcher2012,moses2015saturnbeacon,cavalie2015saturnbeacon}. This shows how the troposphere can potentially influence the stratosphere, and we seek to investigate whether similar processes might be at work on Uranus.

Breakthroughs in ice giant mid-infrared science were once dominated by Voyager IRIS data, taken from the 1986 flyby during Uranus' southern summer solstice \citep[e.g.,][]{hanel1986,flasar1987,orton2015}. Voyager 2 is still the only spacecraft to have visited the ice giants and therefore the only source of close-up remote sensing. The latter data showed meridional differences in the upper tropospheric (80 mbar) temperature structure \citep{conrath1998}, with cool mid-latitudes compared to a warmer equator and poles. Ground-based observations over the past decade have shown that this temperature contrast has not changed significantly over time, despite the dramatic change in season \citep{orton2015, roman2020}.

At present, observations are mostly obtained using ground-based telescope facilities, but they have limitations in the mid-infrared due to telluric absorption at critical wavelengths. It is then unsurprising that the Spitzer Space Telescope, especially the Infrared Spectrometer (IRS) instrument, has provided a significant leap in mid-infrared observation capabilities with its sensitivity being two orders of magnitude greater than previous missions \citep{houck2004infrared}. This high sensitivity has allowed us to go beyond the 17-24 $\mu$m ground-based data, and the 25-50 $\mu$m Voyager data, to explore the individual gaseous features that include mid-infrared wavelengths at short as 5 $\mu$m. Alongside previously determined abundances of methane and acetylene \citep[][etc.]{orton1987,encrenaz1998}, \cite{burgdorf2006} obtained the first clear detections on Uranus of ethane, propyne, diacetylene and carbon dioxide in this wavelength range using data from the Spitzer/IRS acquired in 2004.

The Spitzer IRS has been used to observe both of the ice giants multiple times. In this paper we use the archival data from December 2007 to analyse a temporal variation and therefore longitudinal variation across the spectra of Uranus. This variability as the planet rotates is in the emission of hydrocarbon species observed in the stratosphere. \cite{orton2014mid1,orton2014mid2} conducted a globally averaged study using this same data set. They determined the globally-averaged mean temperature structure and composition of the upper troposphere and stratosphere. Their derived stratospheric temperatures were compatible with those from Voyager 2 UV occultations \citep{herbert1987} but were higher than those derived from Voyager 2 radio occultations between around 1 and 10 mbar \citep{lindal1987,sromovsky2011}. They determined the abundances of acetylene, ethane, propyne, diacetylene, carbon dioxide and tentatively methyl. They provided no obvious evidence for changes in these hydrocarbon abundances with time from 1986 to 2007. The derived acetylene profile was, however, much greater than that determined by \cite{bishop1990} using Voyager UVS occultations and was postulated to be a potential physical increase over time.

Spatially-resolved ground-based imaging of Uranus in the mid-infrared has revealed that emission from stratospheric acetylene is relatively enhanced at mid and high latitudes compared to that at the equator \citep{roman2020}. \cite{roman2020} found these spatial differences to be consistent with either a 16-K latitudinal gradient in the stratospheric temperatures or a factor of 10 gradient in the stratospheric acetylene abundance, arguing in favor of the latter based on the vertical motions implied by complementary upper-tropospheric observations. Until these observations in 2009 and 2018, there had been no mid-infrared measurements of spatial structure in the stratosphere of Uranus, and these will be used to assess the sources of longitudinal variability observed by Spitzer.

In section \ref{S:obs} we discuss how the observations were acquired and reduced to provide spectra, the calibration of these new reductions and compare them to those reduced by \cite{orton2014mid1}. In section \ref{S:var} we analyse the longitudinal variation in the equinox data. In section \ref{S:epoch} we compare the main 2007 dataset to those taken in 2004 and 2005 which also have been newly reduced. In section \ref{S:method} the methodology that has been used to perform the retrievals is outlined with the results in section \ref{S:results}. We retrieve the globally-averaged temperature and composition from both the high- and low-resolution spectra, along with the separate longitudes, in order to determine whether temperature changes or compositional changes are the most likely cause behind the observed longitudinal variation. We discuss these causes and compounding evidence for them in section \ref{S:disc} as well as a comparison of our global average retrieval results to those obtained in \cite{orton2014mid1,orton2014mid2}. Our findings are compared to spatially-resolved observations at both similar wavelengths \citep{roman2020} and similar observation times \citep{sromovsky2009,depater2011,depater2013} to understand the source of the variability. 

\section{Observations}
\label{S:obs}

Due to Uranus' extremely high obliquity we can only observe both northern and southern hemispheres simultaneously when the planet is close to its equinox. The northern spring equinox occurred on the 7th December 2007 with the aforementioned Spitzer observations occurring just 10 days later. NASA's Spitzer \citep{werner2004spitzer} Infrared Spectrometer (IRS) \citep{houck2004infrared} used four modules of varying resolution and wavelength range between 5.23 and 37.2 $\mu$m. The two low-resolution modules, short-low (SL) and long-low (LL), are both split into two separate sub-slits and have three orders: two main orders and a bonus third order overlapping regions between the two. The two high-resolution modules, short-high (SH) and long-high (LH), have ten orders each between 9.9 and 37.2 $\mu$m. The slit sizes (width x length) of modules SL, LL, SH and LH are 3.7'' x 57'', 10.6'' x 168'', 4.7'' x 11.3'' and 11.1'' x 22.3'' respectively. Further details of these modules can be found in \cite{orton2014mid1} Table 1. 

The 3.35-arcsec angular diameter disc of Uranus is treated as an unresolved source and therefore the resulting spectra are disc-averaged. The Spitzer data have been re-analysed using the most up-to-date pipeline software available from NASA's Spitzer Science Centre (SSC) resulting in minor changes over the previous reduction by \cite{orton2014mid1} that are detailed in section \ref{SS:errs}. 

%%TABLE
\begin{table*}[ht!]
\centering
\small
\begin{tabular}{l l l l l l}
\hline
\textbf{Instrument} & \textbf{Date(s)} & \textbf{Modules/Filters} & \textbf{Wavelength(s) $\lambda$ ($\mu$m)} & \textbf{$n_{long}$} & \textbf{References} \\
\hline
Spitzer IRS & 2007-12-16 - 17 & SL,LL2,SH,LH & 5.23-37.20 & 4 & \cite{orton2014mid1,orton2014mid2} \\
{} & 2005-07-06 - 07 & SL,SH,LH & 5.23-37.20 & 4 & - \\
{} & 2004-11-12 - 13 & SL,SH,LH & 5.23-37.20 & 3 & \cite{burgdorf2006} \\
\hline
Keck II NIRC2 & 2007-12-12 - 13 & H & 1.6 & 4 & \cite{depater2011} \\
{} & 2007-12-12 - 13 & K' & 2.1 & 4 & \cite{depater2013} \\
\hline
VLT-VISIR & 2009-08-05 - 06 & NeII 2 & 13.04 & 2 & \cite{roman2020}) \\
\hline
\end{tabular}
\caption{A summary of observations of Uranus from Spitzer-IRS, Keck II-NIRC2 and VLT-VISIR referred to in this investigation. $n_{long}$ is the number of distinct central meridian longitudes observed. Spitzer data were downloaded from the Spitzer Heritage Archive (\url{https://sha.ipac.caltech.edu/applications/Spitzer/SHA/}) and reduced and stored on Github (\url{https://doi.org/10.5281/zenodo.4617490}).}
\label{tab:obssum}
\end{table*}

Table \ref{tab:obssum} summarises the observations of Uranus made with Spitzer for use in this and future longitude studies of the planet. The Keck and VLT observations referred to in section \ref{S:disc} are also included.

\subsection{Observation Times and Longitude Calculations}

Table \ref{tab:modtime} lists the observation times and associated longitudes for the 2007 data and is an update on that reported by \cite{orton2014mid1}. The IRS's modules are in separate, fixed boxes with no moving parts, so that only one instrument module can be exposed at a time. Switching apertures involves the re-pointing of the entire telescope using a three-axis stabilized control system made up of four reaction wheels \citep{werner2004spitzer}. The time this takes causes a spread of data points across the multiple longitudes. Each module observed Uranus during four separate longitude campaigns, giving a good spread of data across the disc. It should be noted that each spectrum taken at each longitude refers to the disc that is centred on that and the latitude 2.69\degree South for all observations. This is the location of the sub-Spitzer point. The IAU2009 longitude is based on the `System III' prime meridian rotation angle of the magnetic field assuming a rotation period of 17.24 hr. Pole direction (thus latitude) is relative to the body dynamical equator (\href{https://ssd.jpl.nasa.gov/horizons.cgi}{JPL Horizons On-line Ephemeris System}).

%%TABLE
\begin{table*}[ht!]
\centering
\small
\begin{tabular}{l l l l l l}
\hline
\textbf{Module} & \textbf{Longitude} & \textbf{Date} & \textbf{Start Time} & \textbf{End Time} & \textbf{Mean Longitude (\degree E)} \\
\hline
SL & 1 & 2007-12-16 & 17:40 & 18:18 & 209.45 \\
SH & 1 & 2007-12-16 & 18:18 & 19:33 & 229.11 \\
LL & 1 & 2007-12-16 & 19:34 & 19:38 & 243.20 \\
LH & 1 & 2007-12-16 & 19:42 & 19:44 & 245.64 \\
SL & 2 & 2007-12-17 & 00:06 & 00:43 & 343.60 \\
SH & 2 & 2007-12-17 & 00:44 & 01:59 & 3.43 \\
LL & 2 & 2007-12-17 & 02:00 & 02:07 & 18.05 \\
LH & 2 & 2007-12-17 & 02:08 & 02:10 & 19.97 \\
SL & 3 & 2007-12-17 & 06:46 & 07:24 & 122.97 \\
SH & 3 & 2007-12-17 & 07:24 & 08:39 & 142.63 \\
LL & 3 & 2007-12-17 & 08:40 & 08:47 & 157.25 \\
LH & 3 & 2007-12-17 & 08:48 & 08:50 & 159.16 \\
SL & 4 & 2007-12-17 & 13:15 & 13:52 & 258.17 \\
SH & 4 & 2007-12-17 & 13:53 & 15:08 & 278.00 \\
LL & 4 & 2007-12-17 & 15:09 & 15:16 & 292.62 \\
LH & 4 & 2007-12-17 & 15:17 & 15:19 & 294.53 \\
\hline
\end{tabular}
\caption{Spitzer-IRS 2007 campaign observing times (UT) and disc-centred longitudes (\degree East, system III) that are associated with each module as calculated by JPL Horizons with a 17.24 hr rotation period and sub-Spitzer point located at 2.69\degree S.}
\label{tab:modtime}
\end{table*}

\begin{figure}[ht!]
\centering\includegraphics[width=1\linewidth]{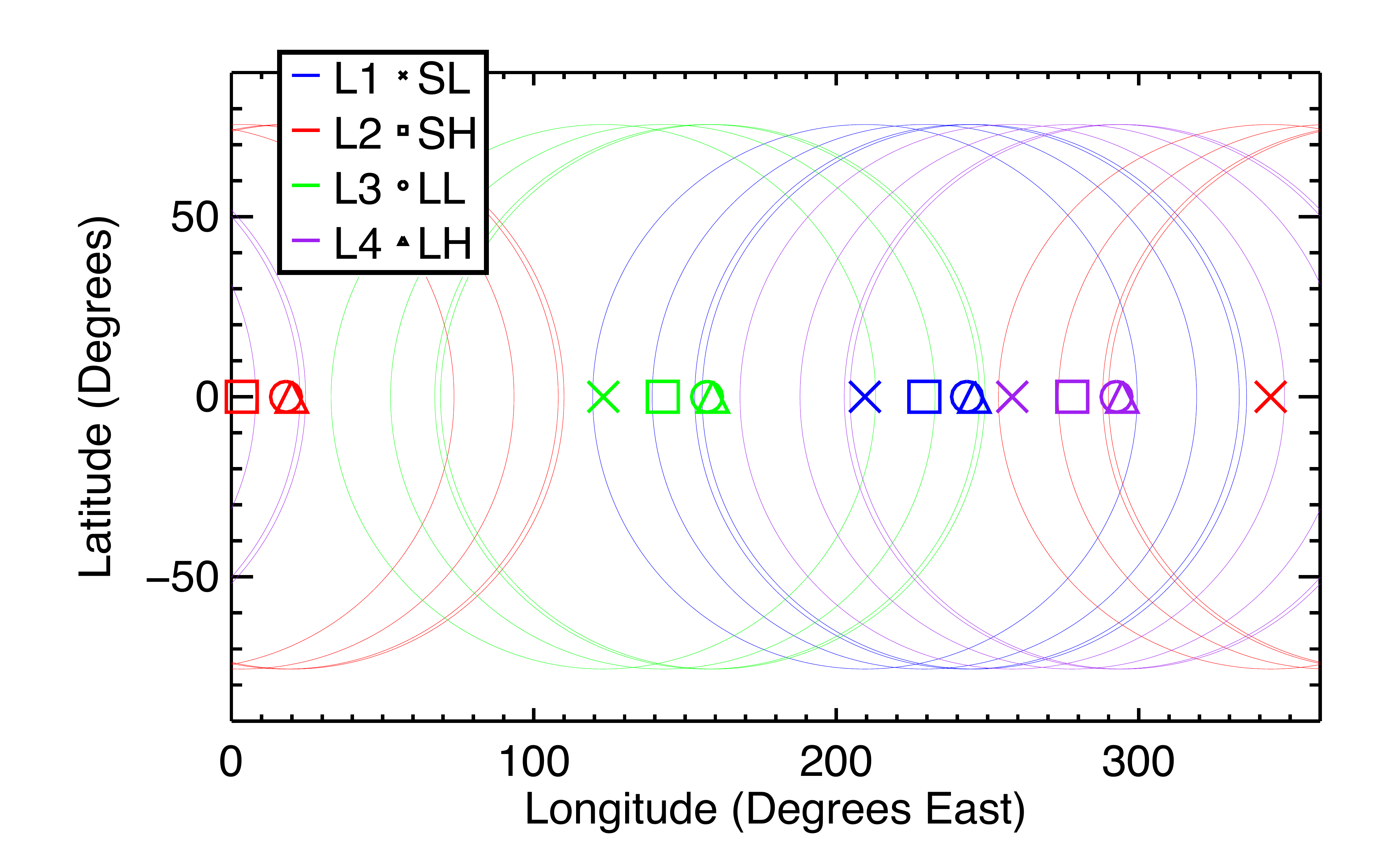}
\caption{The Spitzer exposure ranges for all modules across all longitudes of Uranus for the 2007 campaign. Each longitude is displayed in a unique colour and each module is shown by a unique symbol. Relative latitude is centred on 2.69\degree South.}
\label{fig:hemex}
\end{figure}

The longitudes in Table \ref{tab:modtime} are shown visually in Figure \ref{fig:hemex}. The four separate longitude campaigns are labelled 1 to 4. The circles around each mean longitude point illustrate the disc of the planet and how the exposed hemispheres overlap with the neighboring longitudes.  Longitude 1 and 4 are taken only around $50\degree$ apart but after one full rotation of the planet.

\subsection{Spectral Extraction}

The reduction of the data was done using the software available from the SSC (\url{https://irsa.ipac.caltech.edu/data/SPITZER/docs/dataanalysistools/}), following the steps of \cite{orton2014mid1} with only minor changes detailed below. Most of the available software has received updates and are now the final versions that were not available when the data of \cite{orton2014mid1,orton2014mid2} were reduced in 2008. 

The raw form of data that are available from the heritage archive are FITS files of the 2D detector images. Each observation has two nods that are a dither in different parts of the same detector slit. The SSC software, \textit{irsclean\_mask v2.1.1}, was used to mitigate the effects of bad pixels on the detector that come from dark currents and historically dead pixels. The edited masks from the previous 2008 reduction were not available, so additional bad pixels that were not included in the default campaign-specific mask had to be identified by eye. Bad pixels were identified if they were clearly dead or if they exhibited intermittent large changes in brightness over multiple exposures. The multiple exposures for each module and nod were co-added using the provided \textit{coad} software. For the main 2007 data the SL modules have 25 exposures, LL and SH both have 15 exposures and LH has 4 exposures. The sample integration time per exposure is around 1 second for every module apart from SH that has an integration time of 4 seconds. The two nods were kept separate until the very end of the reduction process.

For SL, the sky subtraction uses the fact that when sub-module SL1 is pointed at the source, SL2 is pointed at blank sky, and vice versa. Subtracting them from one another required no other sky-subtraction technique, unlike \cite{orton2014mid1} who also subtracted the sky directly from the opposite end of the slit from the source. This method of subtracting directly from the slit provided negligible changes to the final spectra, so was omitted. For LL, subtraction of the two separate nods was required because LL1 was unavailable due to expected detector saturation. This method was used rather than subtracting directly from the slit due to the source visibly occupying the majority of the length of the slit with no consistent area to subtract from. Subtracting the nods is a proven method for background subtraction \citep{sloan2004}, and it replicates the approach used for the SL modules. SH and LH both had campaign-specific off-source pointings that could be directly subtracted from the on-source pointings.

The SSC-provided SPitzer IRS Custom Extractor (\textit{SPICE v2.5.1}) extracts the spectra from the co-added, cleaned and sky-subtracted detector images. The settings used are the same as in \cite{orton2014mid1} although the version of the software was updated in August 2013. A combination of the SSC contributed software \textit{stinytim v2.0} and methods by \cite{orton2014mid1} accounts for the overfilling of the slit by the disc of Uranus by generating a point-spread function (PSF) for the IRS as a function of wavelength (0.5 $\mu$m increments were used). Unlike in the previous reduction of \cite{orton2014mid1}, we also corrected the LL spectra for this overfilling factor as we found it had a significant effect, with details described in section \ref{SS:errs}. This same method was used for all other modules. SSC software \textit{IRSFRINGE v1.1} was used on the LH data as in \cite{orton2014mid1} to defringe the orders and correct for this effect that originates from the detector substrates.

\begin{figure*}[ht!]
\centering\includegraphics[width=0.8\linewidth]{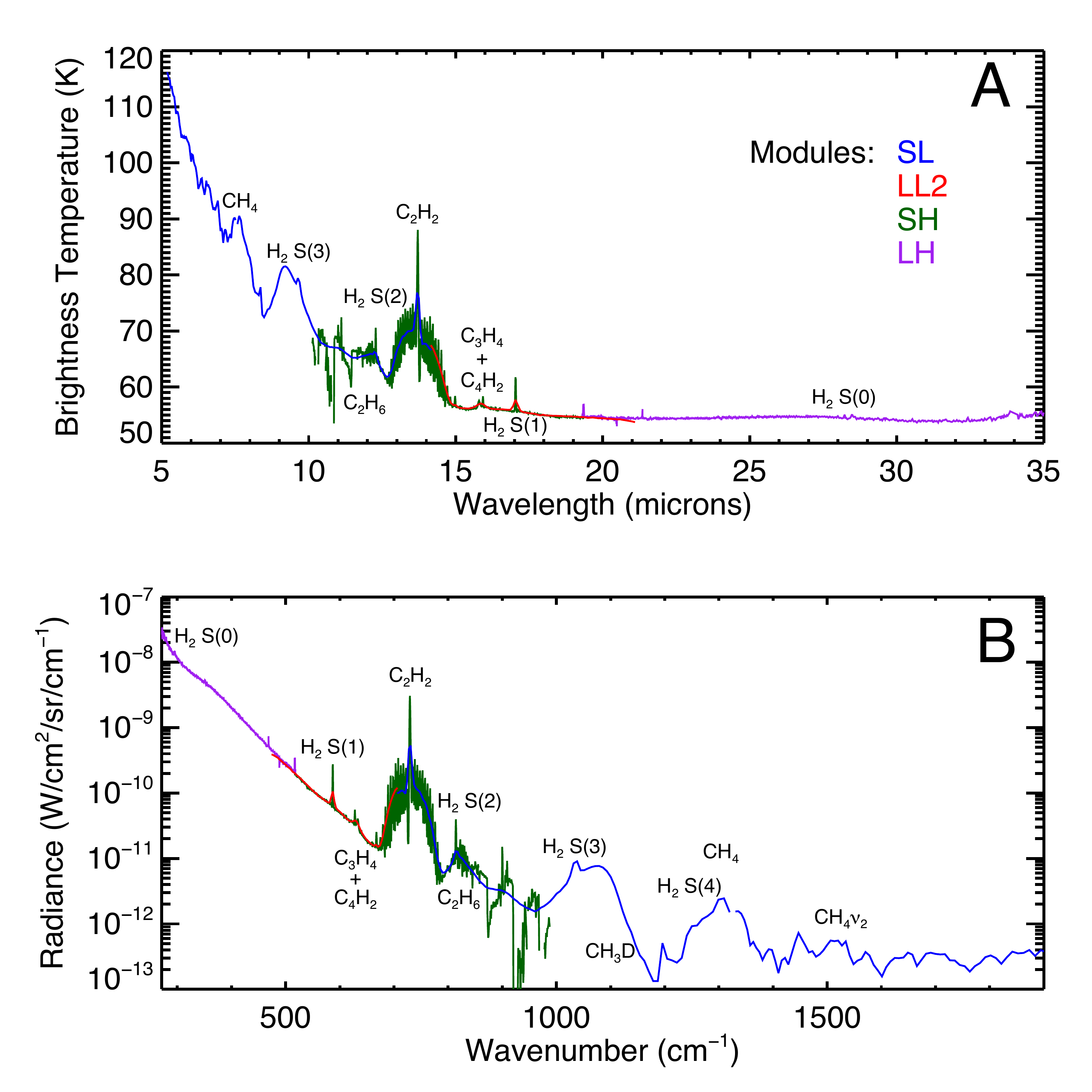}
\caption{The global average spectrum of Uranus for observations made on 16th-17th December 2007 using all available modules of the Spitzer Space Telescope Infrared Spectrometer. The modules are shown in different colours: Short-Low (SL) in blue, Long-Low second order (LL2) in red, Short-High (SH) in green and Long-High (LH) in purple. Panel A shows brightness temperature vs. wavelength and panel B shows radiance vs. wavenumber. Clear spectral features are labelled with their chemical names.}
\label{fig:fullspec}
\end{figure*}

The globally-averaged spectra from this updated reduction pipeline are shown in Fig. \ref{fig:fullspec}. Data are converted from Janskys into radiance and brightness temperature using an angular diameter of Uranus of 3.35 arcsec from the vantage point of Spitzer. The data show significant drop-outs at the edges of each order/sub-module. To compensate for this, two spectral points from either end of the low-resolution data were omitted from subsequent analysis. Ten points were omitted from high resolution data.

The spectra from the SH modules showed a slope that disagreed with the better-calibrated SL and LL data, so they were pivoted (i.e. corrected with a linear dependence on wavelength) and scaled to match at the same wavelengths using the same method as \cite{orton2014mid1}. Due to the small changes in the data from the updated reduction pipeline the pivoting values were slightly changed. The scaling factors used on each order were unchanged from the previous reduction but the pivoting factors required small changes. The need to scale up the SH data to the low-resolution modules by around 5\% is also highlighted in \cite{burgdorf2006} for the 2004 data.

\subsection{Errors and Calibration}
\label{SS:errs}

The measurement errors provided in the reduced data come from the Spitzer pipeline and are calculated in SPICE as a 1-sigma error. These errors do not take into account the reproducibility of the measurements so these were combined manually. For SL2 the average 1-sigma error from the pipeline is 6\%, for SL1 and LL2 it is 1\%. Additional error is added to account for the correction done by \textit{stinytim}. This is estimated as the average correction factor for each module, where both SL modules are 7\% and LL is 3\%. These module-specific uncertainties are combined in quadrature with the 4\% and 2\% errors due to disagreements with standard stars and differences in the infrared calibration of the telescope respectively \citep{orton2014mid1}. This yields a total radiance uncertainty of 6\% for SL1, 7\% for SL2 and 4\% for LL2. Each percentage error mentioned is the percentage of the measured radiance at every spectral point.

For the high-resolution modules the errors are not so easily derived due to the fact that the SH module data are pivoted and scaled to the low-resolution data and fringes are corrected for in the LH module. By using the same methodology as the low-resolution with these additional errors added in quadrature we estimate an error of 9\% for SH and 5\% for LH.

\begin{figure}[ht!]
\centering\includegraphics[width=1\linewidth]{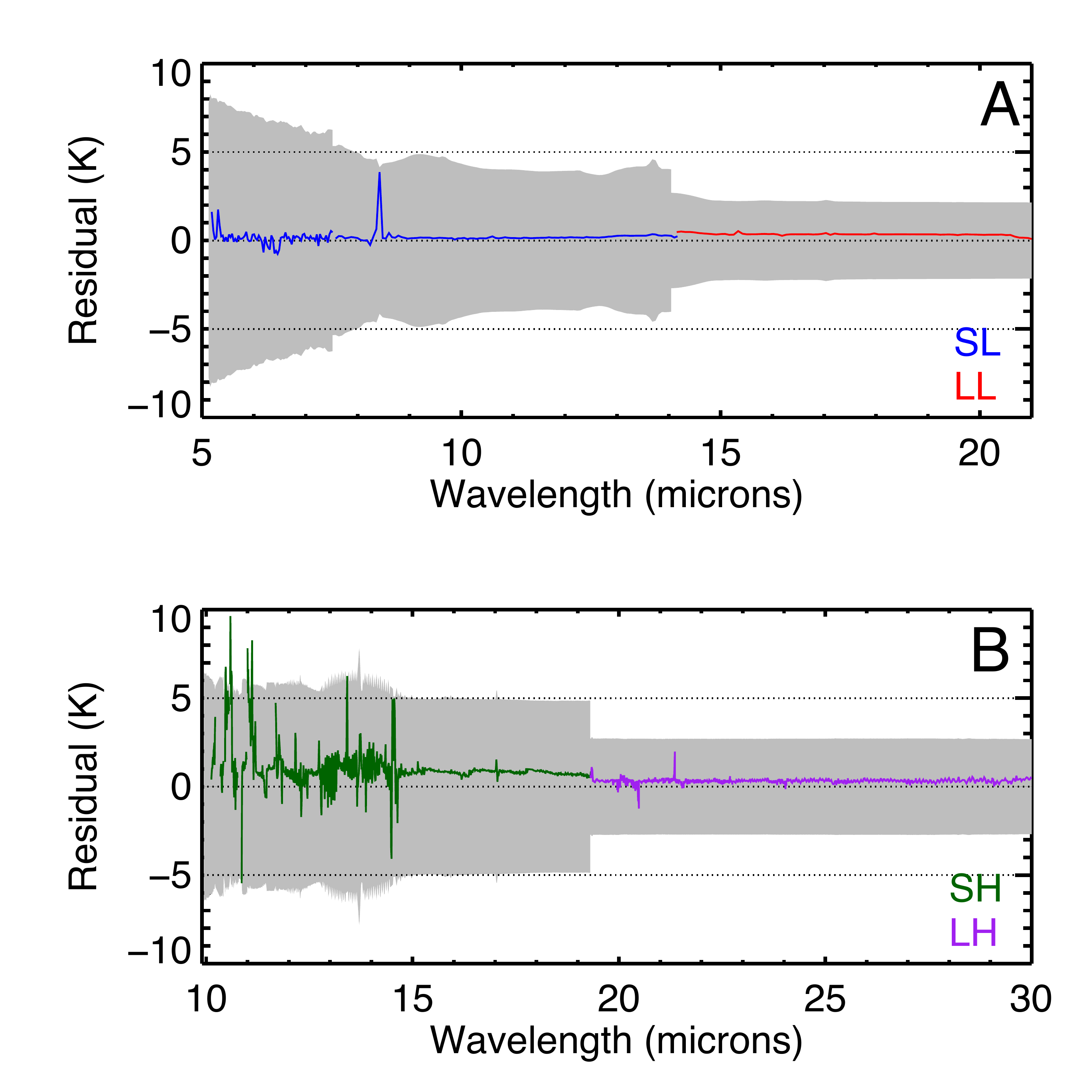}
\caption{The difference between this work and the previously reduced data by \cite{orton2014mid1} with low-resolution modules shown in panel A and high-resolution modules shown in panel B. The total flux uncertainties associated with the reduction are shown by the grey shaded regions (6\% for SL1, 7\% for SL2, 4\% for LL2, 9\% for SH, and 5\% for LH, converted to brightness temperature).}
\label{fig:compspec}
\end{figure}

Fig. \ref{fig:compspec} compares the previously reduced data from \cite{orton2014mid1} and provides the difference between the two in terms of brightness temperature, showing that the difference is within the total flux uncertainty of the reduction process in most regions. The entire spectrum exhibits a small upward shift (up to 0.3 K in the low-resolution and up to 1 K in the high-resolution) showing that the new reductions are slightly brighter than the old ones. For the LL module, this could be due to the corrections caused by the use of the \textit{stinytim} tool. Otherwise, the extraction of a consistently brighter spectrum is most likely due to changes in the updated SSC tools and products. As previously mentioned, the factor used to scale the high-resolution data to the low was the same as used in \cite{orton2014mid1} so the relative change across the modules has been consistent. This supports the idea that this brighter spectrum is a real change in the data due to improved calibration software and not an error that has propagated through the pipeline.

The bonus order (SL3) between 7.9-8.5 $\mu$m (1176 - 1266 cm$^{-1}$) was inconsistent with its neighboring orders and was not used, consistent with the previous investigation \citep{orton2014mid1}. The large difference seen at around 8.5 $\mu$m is in a very low radiance region that exhibits a lot of noise. This region has also been omitted from subsequent analysis.

\cite{orton2014mid1} found that the LL flux is 9\% higher than the SL1 flux. We found that the difference based upon this new reduction is around 15\%. This is due to a combination of two factors: (i) The known discontinuity between the two Spitzer modules that has a standard deviation of around 9\% \cite{sloan2004}; and (ii) the 34-degree change in longitude caused by the time difference between each measurement (Table \ref{tab:modtime}). This longitudinal change in brightness is the main focus of this investigation.

The LH module data are well within the errors shown. The 10 - 12 $\mu$m region of the SH module has been omitted from subsequent analysis due to noise. Some of the least consistent features have been cut out from the acetylene and ethane region between 12 and 15 $\mu$m. The high-resolution data have been used as a comparison but the most important scientific conclusions are based only on the low-resolution data due to their consistency and reliability.

\section{Longitudinal Variation}
\label{S:var}

Due to the wide and even spread of longitudes (Table \ref{tab:modtime}, Fig. \ref{fig:hemex}) we are able to look at the variation in 2007 in detail. The spectrum contains several different complex hydrocarbons associated with the lower- and mid-stratosphere \citep{moses2018} and five hydrogen quadrupole lines sensing the upper stratosphere \citep{orton2014mid1}. The hydrogen-helium collision-induced absorption (CIA) is a main feature of the spectrum present at the higher pressures of the troposphere, creating the overall shape of the entire spectrum \citep{orton1986}. Monodeuterated methane (CH$_3$D) on Uranus is also associated with these higher pressures due to its low abundance, unlike CH$_4$ that senses the lower-pressure regions of the stratosphere. Complex hydrocarbons that are the products of methane photodissociation to be investigated include ethane (C$_2$H$_6$), three ro-vibrational bands of acetylene (C$_2$H$_2$), diacetylene (C$_4$H$_2$) and methylacetylene a.k.a. propyne (CH$_3$C$_2$H a.k.a. C$_3$H$_4$).

\begin{figure*}[ht!]
\centering\includegraphics[width=0.8\linewidth]{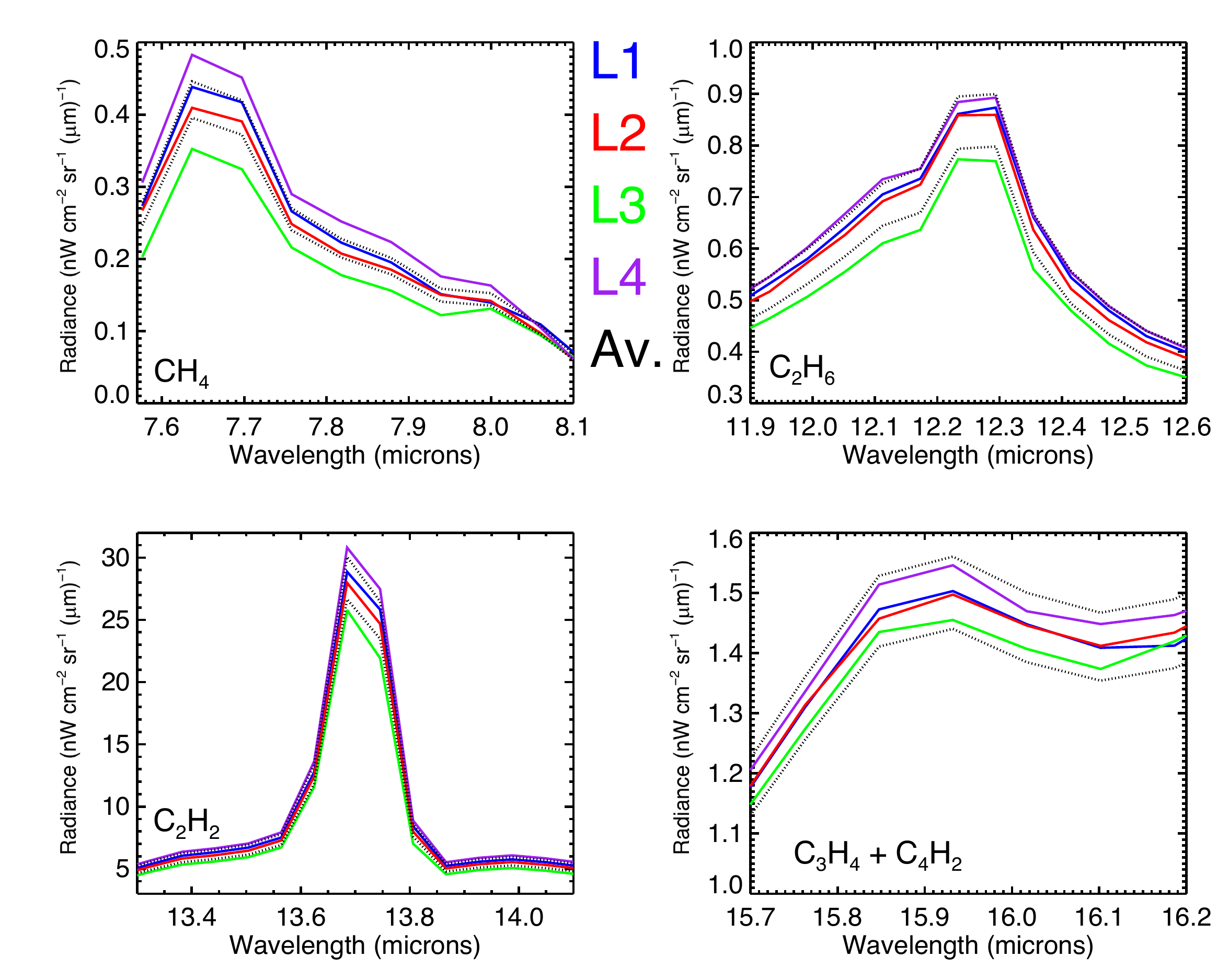}
\caption{The radiances of the four different longitudes in the main hydrocarbon emission features for the low-resolution data from Spitzer-IRS in 2007. Longitude 1 to 4 are shown as blue, red, green and purple lines respectively. The upper and lower limits (based on the measurement errors stated in section \ref{SS:errs}) of the global-average radiance are shown as black dotted lines. The radiance is on a linear scale for direct comparison and features the same colour scheme as Fig. \ref{fig:hemex} for comparison.}
\label{fig:seplong}
\end{figure*}

\begin{figure}[ht!]
\centering\includegraphics[width=1\linewidth]{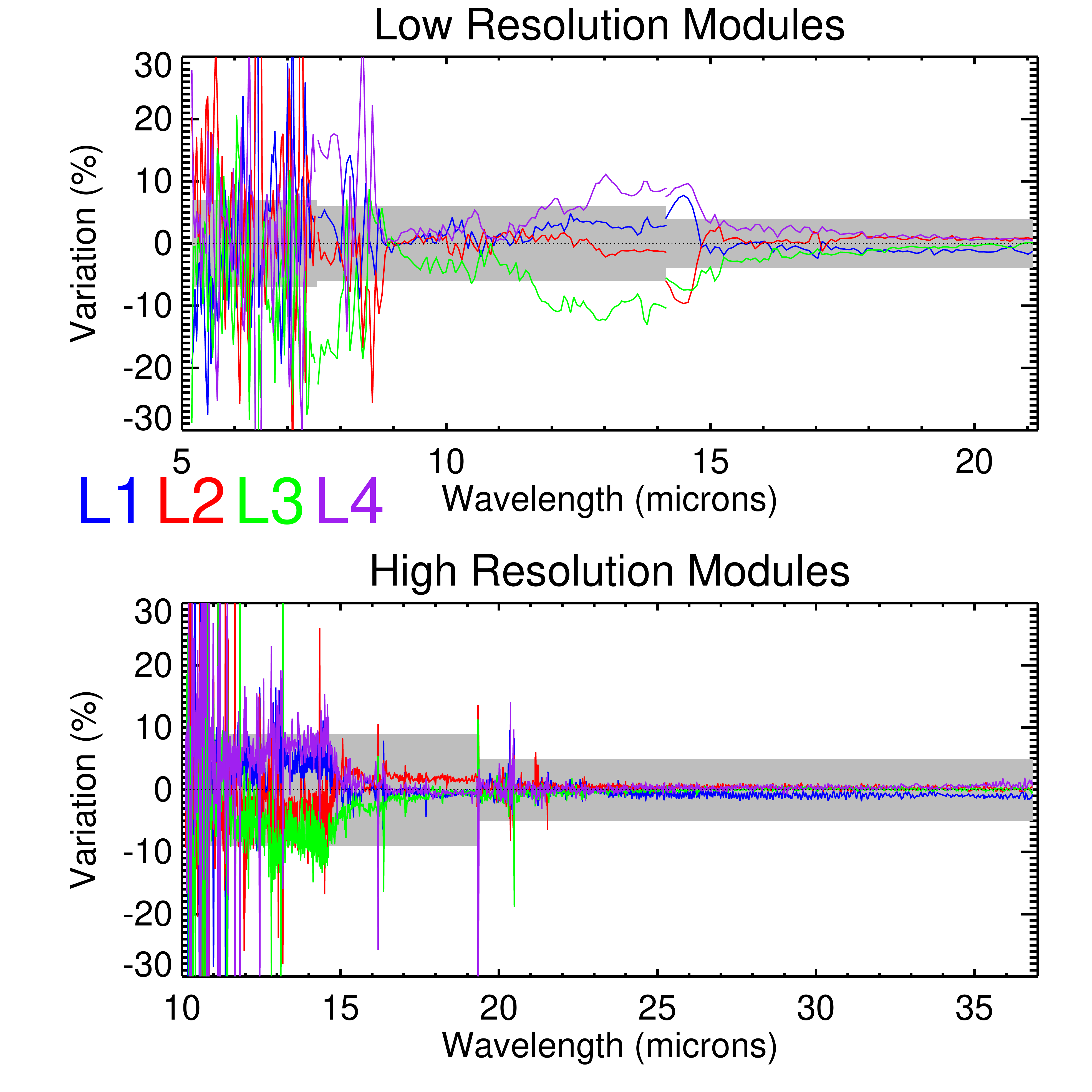}
\caption{The percentage difference in radiance between the four separate longitudes and the global average across both the low-resolution modules (top panel) and high-resolution modules (bottom panel). Measurement uncertainties for each module are shown as the grey shaded regions.}
\label{fig:longdiff}
\end{figure}

The four intervals in Figure \ref{fig:seplong} show a clear pattern emerging between the four separate longitudes. Longitude 3 is significantly dimmer than longitude 4 in every interval. These two longitude observations were taken 135 degrees apart with roughly 6.5 hours elapsing between observation mid-times. Longitude 1 and 2 show very similar radiance values despite nearly equivalent separations in longitude and time. In ranges of the spectrum where a variation can be seen, we see this same pattern across the low and high resolution spectra (Fig. \ref{fig:longdiff}). The acetylene region around 13 $\mu$m and the methane region around 8 $\mu$m show this pattern the most clearly above uncertainty values.

\begin{figure*}[ht!]
\centering\includegraphics[width=0.8\linewidth]{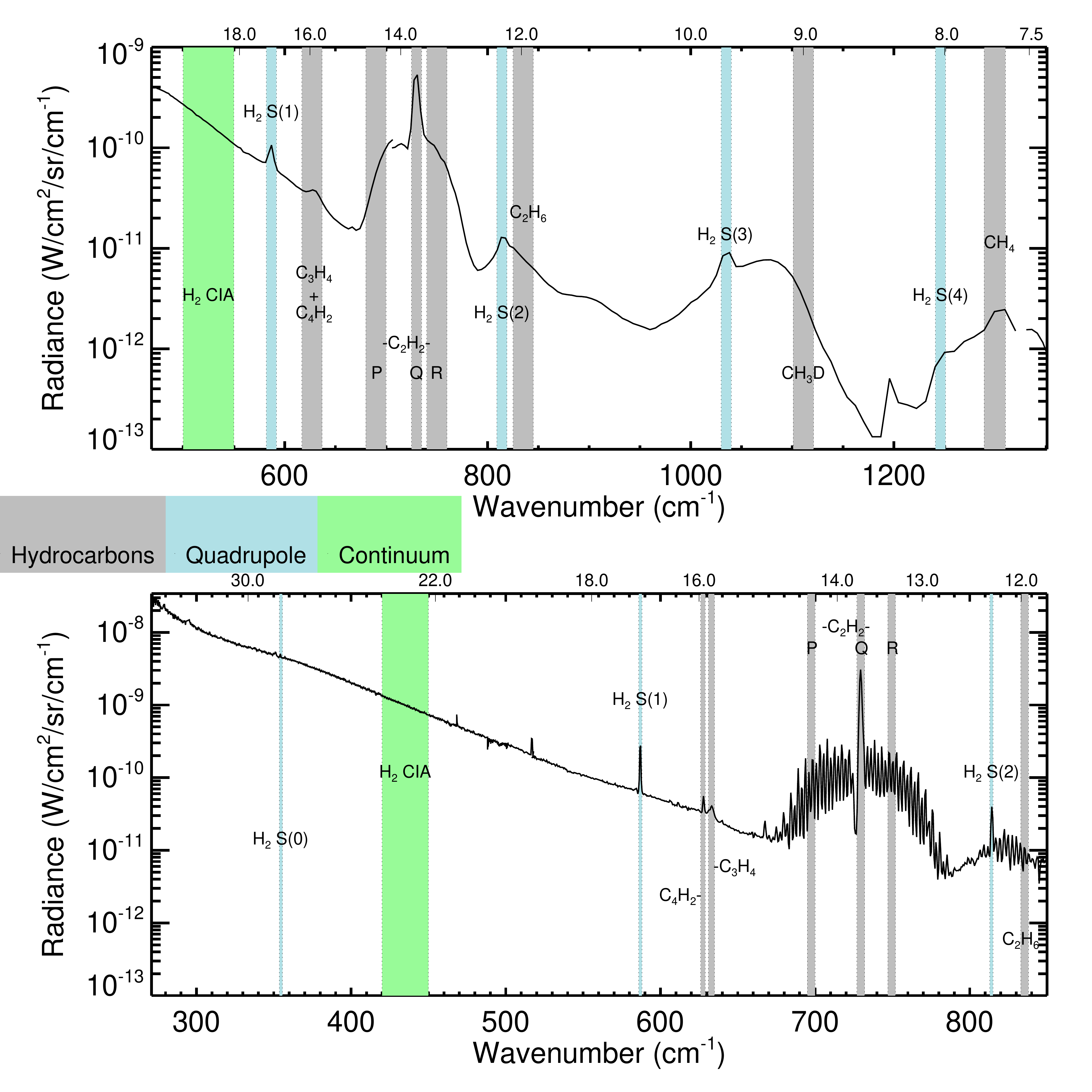}
\caption{Wavenumber intervals selected for different species in longitudinal variation calculations. The global-average low-resolution data are shown in the top panel and the global-average high-resolution data in the bottom panel. Wavelength values shown on top abscissa of both graphs in microns.}
\label{fig:specints}
\end{figure*}

\begin{figure*}[ht!]
\centering\includegraphics[width=0.8\linewidth]{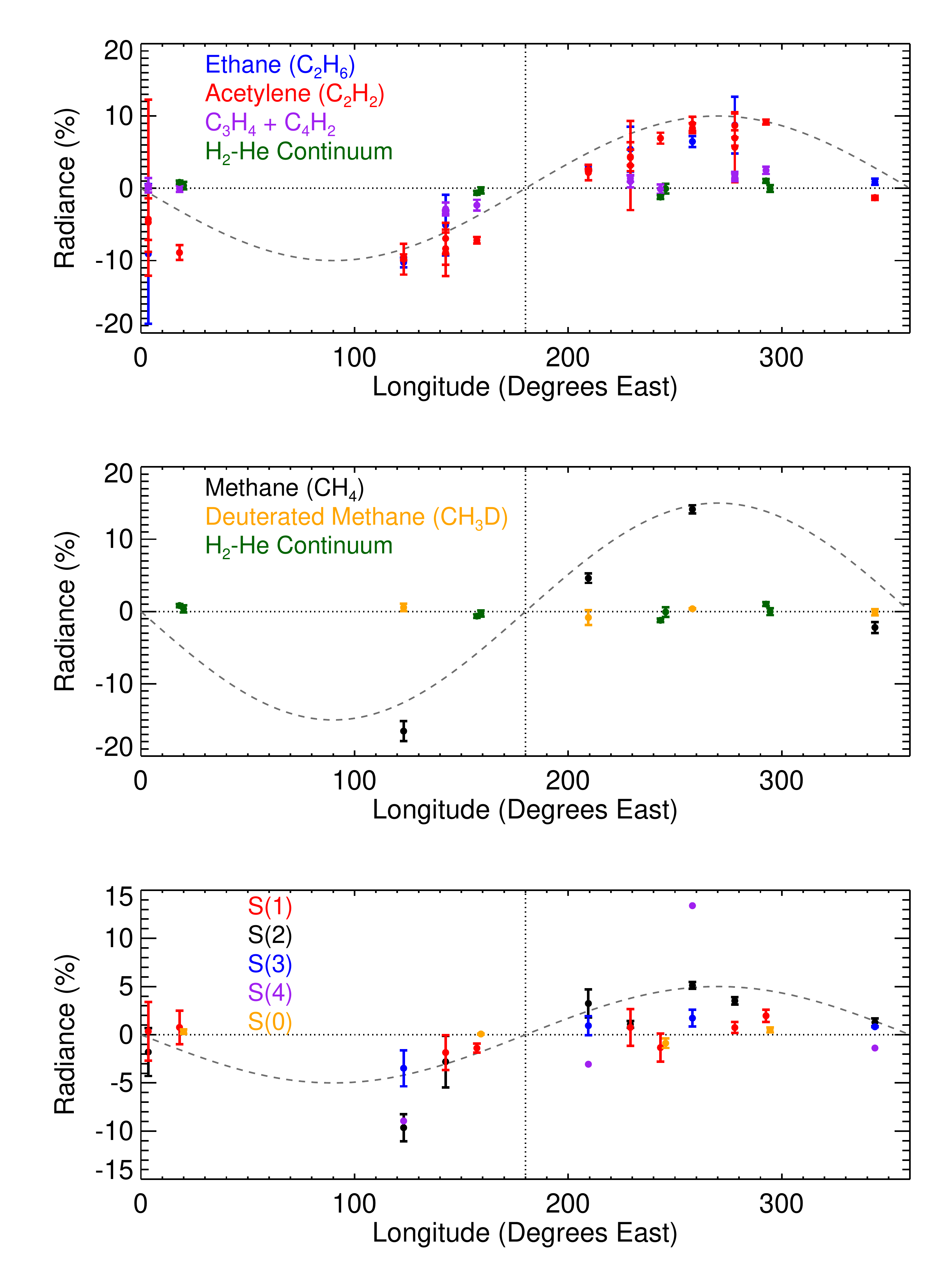}
\caption{The percentage radiance difference from the global average of chemical species across $360\degree$ of Uranus from all four Spitzer modules. Non-negligible standard errors are displayed not including measurement errors for clarity. Top panel: Complex hydrocarbon species and the hydrogen-helium continuum with 10\% amplitude wavenumber-1 sinusoid displayed for reference. Middle panel: Methane variants and the hydrogen-helium continuum with 15\% amplitude wavenumber-1 sinusoid. Bottom panel: hydrogen quadrupole species with 5\% amplitude wavenumber-1 sinusoid displayed. It is merely coincidental that the sinusoidal pattern of the data has a phase-shift of zero, this has not been forced.}
\label{fig:longvary}
\end{figure*}

We assess the variations in discrete channels sensitive to the different emission features in Fig. \ref{fig:specints}. The radiances within each interval are averaged and compared to the global-average radiance to get a percentage radiance difference for each species at each observed longitude (Fig. \ref{fig:longvary}). The intervals shown in Fig. \ref{fig:specints} were selected using known central wavelengths for each species at Uranus. The hydrocarbons were taken from GEISA-2003 compilation \citep{jacquinet2003geisa}. We use the locations and widths of the hydrogen quadrupole lines from \cite{fouchet2003,orton2014mid1,fletcher2018}. The widths over which to average were deduced from the widths of the features (slightly different for high and low resolutions). The hydrogen-helium CIA is present throughout the spectrum but was chosen to be represented by the wide ranges indicated because of a lack of other features in these regions. We chose two different wavenumber ranges for each resolution to gain two independent values of the continuum variation for analysis. 

The complex hydrocarbons in the top panel of Figure \ref{fig:longvary} appear to vary from the mean in a noticeably sinusoidal fashion. There is a difference in radiance of up to 20\% between the variations from $0\degree$ to $180\degree$ compared with those between $180\degree$ and $360\degree$. The most variation occurs in acetylene with ethane also showing the same pattern. Diacetylene and propyne and the 16 $\mu$m region that encompasses both species in the low resolution observations show a small amount of variation that agrees with the overall trend of around 2 - 3\%. The collision-induced continuum is also plotted for comparison, showing variations smaller than 1\%.

With stratospheric methane emission showing the largest variation of up to 15\% it is interesting to see it contrasted to the tropospheric species in the middle panel of Figure \ref{fig:longvary}. The emission as a function of longitude from the hydrogen-helium continuum is relatively constant in both the high and low resolution bands, varying by less than 1\%. Deuterated methane displays a variation that is similarly negligible ($<$ 2\%), suggesting that this variation is only present in the stratosphere.

The different features of the hydrogen quadrupole are associated with different altitudes. S(1) has its maximum contribution near 0.1 mbar. S(2) and S(3) have maximum contributions at lower pressures of around 1 $\mu$bar. The bottom panel of Figure \ref{fig:longvary} shows a 4-5\% variation in these lower-pressure quadrupole lines than those sensing higher pressures. The S(4) quadrupole line shows the largest variation (10-13\%) but this is most likely due to it being embedded in the highly variable methane feature. S(0) has only a very weak stratospheric contribution at its central peak of 28.2 $\mu$m (354 $cm^{-1}$) \cite{fletcher2018}. Here it primarily senses the upper troposphere and so we have another confirmation of a lack of variation at these altitudes.

The apparent wavenumber-1 sinusoidal pattern is present in all three panels of Figure \ref{fig:longvary} with a minimum value consistently in the first longitudinal hemisphere and the maximum in the second hemisphere. It must be remembered that each point is just the mid-point of the disc that, in reality, spreads 90\degree east and west from that point as shown visually in Figure \ref{fig:hemex}. Bias in the shape of the variation may be caused by the gap in the longitude spread of data between 18\degree E and 123\degree E. The variation is well above the standard error calculated, shown as error bars. Figure \ref{fig:longdiff} shows the acetylene, ethane and methane ranges also have variations that are above the data measurement errors. There is consistency in the variations between the independently acquired modules in both resolutions. L1 and L4 (the clusters of points between $200\degree$ and $300\degree$) show consistently higher average radiances for the stratospheric species even though these two campaigns were a full rotation apart assuming a rotation period defined by the Voyager magnetic field data (Table \ref{tab:modtime}). This shows evidence of the variability being caused by a localised feature that is stable over short timescales.

\section{2004 and 2005 Data}
\label{S:epoch}

\begin{figure}[ht!]
\centering\includegraphics[width=1\linewidth]{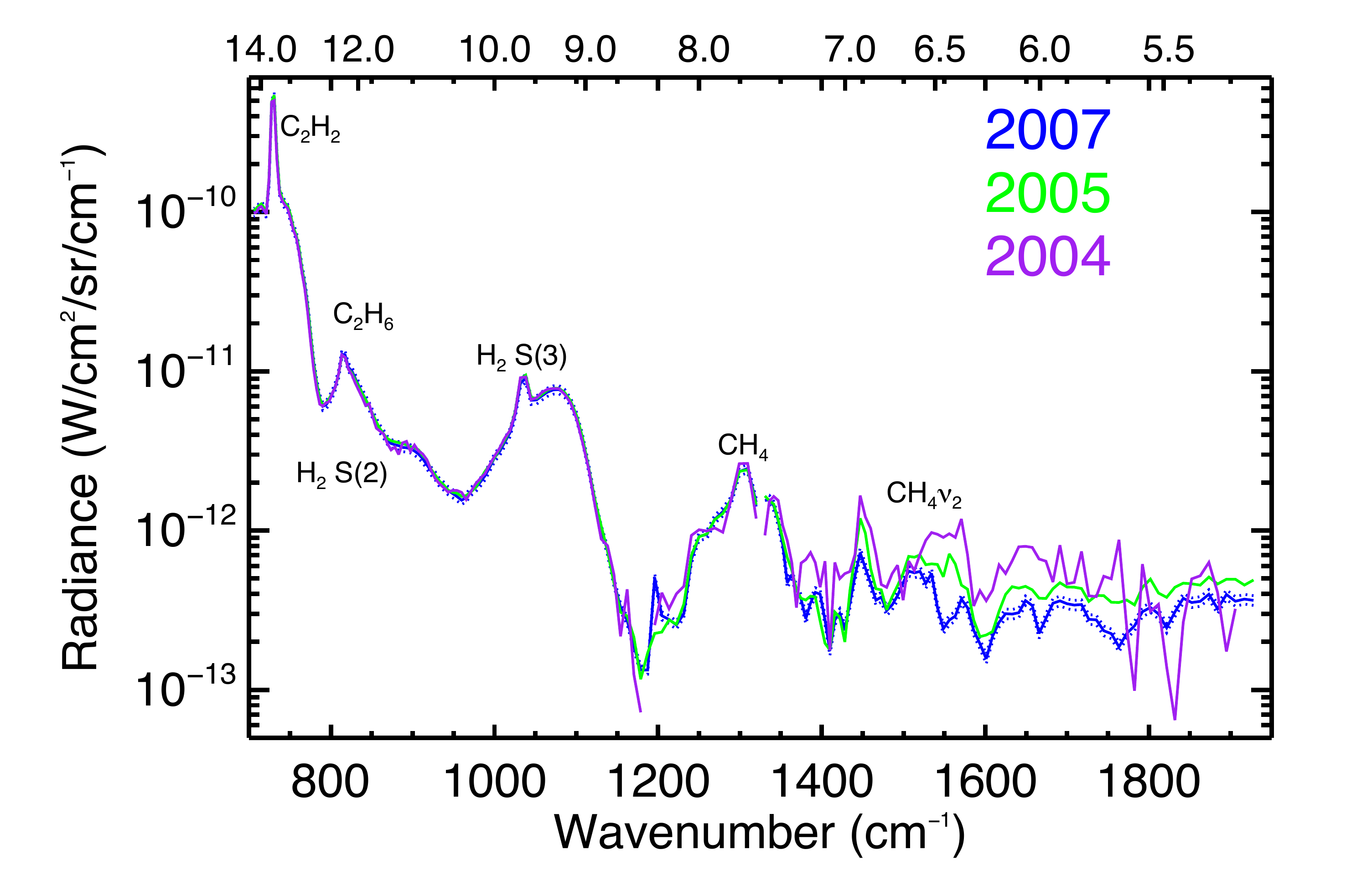}
\caption{The Spitzer-IRS SL module spectra for the three epochs of Uranus (2007, 2005 and 2004) with main features labelled. Measurement error maxima and minima are shown for the 2007 data as dotted lines. Corresponding wavelength units are displayed on top x-axis in microns.}
\label{fig:070504fig}
\end{figure}

We now seek to investigate whether the same variation was visible at different epochs.  Table \ref{tab:obssum} shows that data from July 2005 and November 2004 are also available from the Spitzer Heritage Archive. The 2004 data have been previously published by \cite{burgdorf2006} but the 2005 data remain unpublished. The low resolution data from both epochs have been reduced using the updated pipeline so they can be properly compared. The 2005 data contain four separate longitude observations (Table \ref{tab:0405time}), however the small $6\degree$ separation between longitude 1 and 4 gives us three distinguishable longitudes. The 2004 data have three longitudes that are evenly separated. We have decided to look at only the low-resolution data because of the need to scale and pivot the high-resolution data to the low-resolution results. There are no LL module data for 2004 or 2005 so only the SL module can be compared directly (Fig. \ref{fig:070504fig}).

\begin{table*}[ht!]
\centering
\small
\begin{tabular}{l l l l l l}
\hline
\textbf{Year} & \textbf{Longitude} & \textbf{Date} & \textbf{Start Time} & \textbf{End Time} & \textbf{Mean Longitude (\degree E)} \\
\hline
2005 & 1 & 2005-07-06 & 21:39 & 22:16 & 236.20 \\
{} & 2 & 2005-07-07 & 03:01 & 03:38 & 348.28 \\
{} & 3 & 2005-07-07 & 08:41 & 09:18 & 106.62 \\
{} & 4 & 2005-07-07 & 14:37 & 15:14 & 230.53 \\
\hline
2004 & 1 & 2004-11-12 & 20:03 & 20:07 & 1.87 \\
{} & 2 & 2004-11-13 & 02:06 & 02:10 & 128.19 \\
{} & 3 & 2004-11-13 & 08:12 & 08:17 & 255.73 \\
\hline
\end{tabular}
\caption{The observing time (UT) and disc-centred longitude (\degree East, system III) associated with the SL module of the Uranus data collected in 2005 and 2004 as calculated by JPL Horizons with 17.24 hr rotation period. The sub-Spitzer point is located at latitudes 7.03 - 7.04\degree S for 2005 and 15.22 - 15.23\degree S for 2004.}
\label{tab:0405time}
\end{table*}

\begin{figure}[ht!]
\centering\includegraphics[width=1\linewidth]{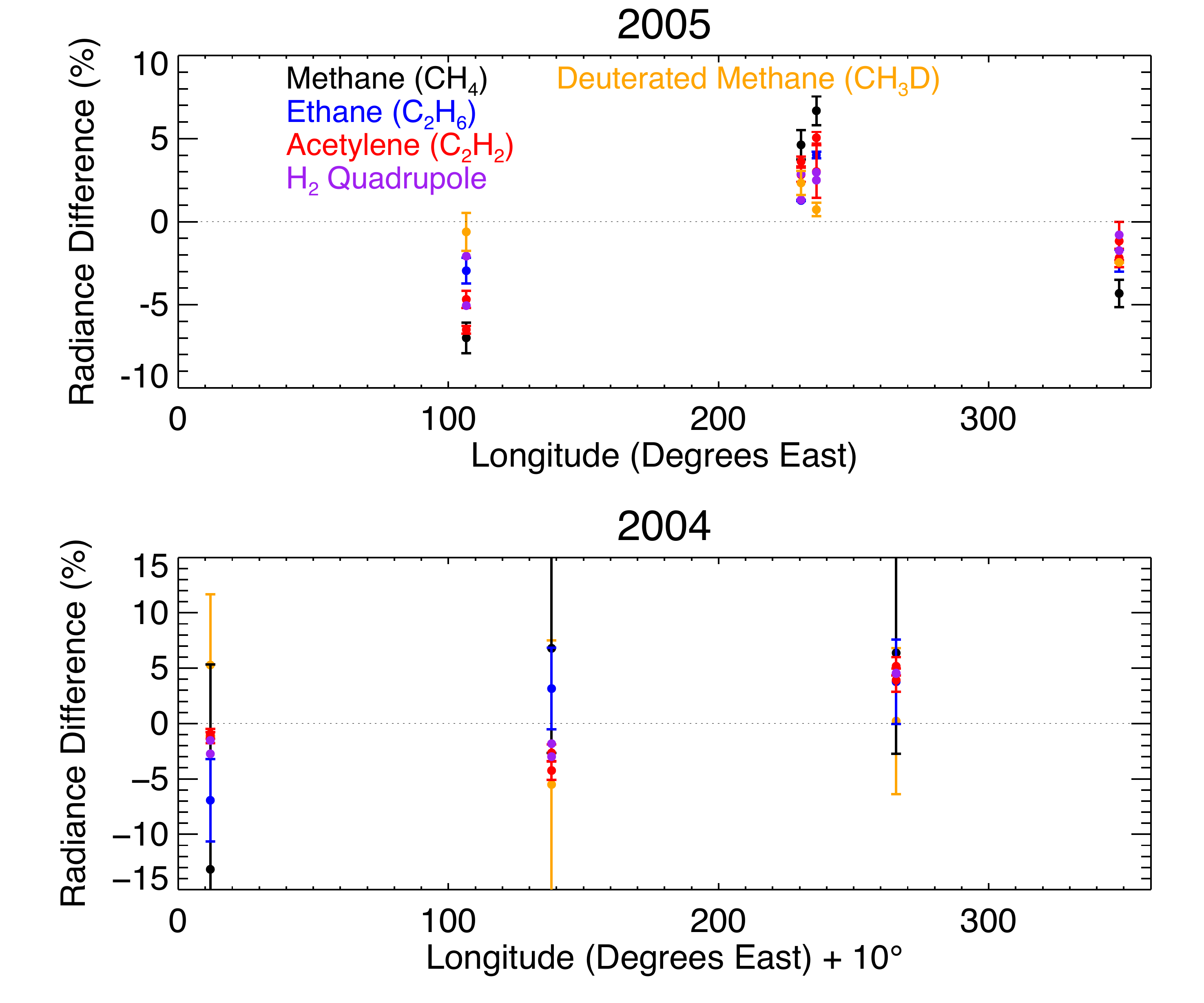}
\caption{The percentage radiance difference from the global average of chemical species across $360\degree$ of Uranus from SL modules of 2005 and 2004 data. Non-negligible standard errors are displayed. The 2004 data has been shifted east by $10\degree$ for clarity.}
\label{fig:var0504}
\end{figure}

The integration times for the 2005 observations are the same as for 2007 with around 37 minutes between the start and end of each exposure. The 2004 data however, only have an integration time of around 4 minutes, so the signal-to-noise ratio is much lower and the resulting spectra less reliable. This can be seen in Figure \ref{fig:070504fig} where the noise at wavenumbers greater than around 1400 cm$^{-1}$ is visibly much more significant than the other two epochs. As an overall trend, Fig. \ref{fig:070504fig} shows a slightly higher radiance for 2005 and then even higher for 2004. This could reflect changes in the visibility of reflective aerosols associated with the southern hemisphere as it approached autumn equinox. The noise can also be seen in the variation of emission where the error bars are much larger, indicating the much larger standard deviation in the data (Fig. \ref{fig:var0504}). This makes drawing any conclusions from the 2004 results difficult.

There is less variation in the 2005 observations over the longitude spread compared to 2007 with changes smaller than 5\%. Longitude 1 (236\degree) and 4 (230\degree) in the 2005 data are only $6\degree$ different but were taken 17 hours apart, after one full rotation of the planet. The data points from the adjacent longitudes in Figure \ref{fig:var0504} are in agreement in their variation from the mean showing that the magnitude and location of the deviation is consistent over short timescales as was observed in 2007. The sparse spread of longitudes make a pattern similar to 2007 impossible to determine.

We have included deuterated methane in the analysis to try to contrast the stratospheric and tropospheric variations. In the 2005 data we see a similar pattern of variation for this species as for the stratospheric ones. Longitude 1 and 4 both show higher radiances compared to longitude 2 (348\degree) and 3 (106\degree) suggesting a variation that is deeper in the atmosphere in 2005 than in 2007. The variation is greatest in methane and acetylene just as it was in 2007. The uncertainties in the 2004 data make this level of analysis non-viable. We conclude that a variation was indeed present in 2005 in excess of the level of noise, although it was smaller than in 2007.

\section{Spectral Retrievals}
\label{S:method}
\subsection{Radiative Transfer Model}

Derivation of the vertical temperature and composition profiles require the use of the optimal estimation retrieval algorithm, NEMESIS, to fit the disc-averaged Spitzer observations \citep{irwin2008nemesis}.  NEMESIS provides two methods of calculating the disc-averaged spectrum - a faster technique using exponential integrals \citep{goodyyung1989} to compute the emission into a hemisphere \citep[used by][for the study of AKARI spectra of Neptune]{fletcher2014neptune}, and a slower but more accurate technique splitting the disc into a number of paths at different emission angles \citep[used by][to analyse Uranus' disc-averaged spectrum from Herschel/SPIRE]{teanby2013uranus}.  We select the latter, more accurate technique for this study.

The goal of this investigation is to develop a consistent retrieval framework for ice giant middle atmospheres. We slowly build up the complexity of the retrieval as we see improvements in the quality of the fits. The data ranges considered in the retrieval for low-resolution spectra are between 7.4 $\mu$m and 21.1 $\mu$m (472 - 1350 cm$^{-1}$) with three parameters (temperature, acetylene and ethane) that have full profile retrievals during the optimisation process. 

For the high-resolution spectra, the range was constrained to 12.0 $\mu$m and 19.0 $\mu$m (526 - 833 cm$^{-1}$) to omit overly noisy ranges without excluding the important features and the addition of the LH data beyond 19 $\mu$m increased the computation time significantly. The following parameters were retrieved as full vertical profiles: temperature, acetylene, ethane and diacetylene. The high-resolution spectra did not cover the methane emission near 7 $\mu$m.  This constraint, combined with the fact that the data were scaled to the low resolution, limited our ability to derive temperatures independently from the high-resolution spectra and was why the low-resolution data were chosen as the main data to analyse.

Our Uranus reference atmosphere has 161 pressure levels and a pressure-temperature profile based on the nominal model from \cite{orton2014mid1} with the assumption that temperatures reach just over 450K at a top pressure of 0.1 $\mu$bar. Alternative temperature profiles, including those mentioned in \cite{orton2014mid1}, were tested and this nominal model was found to provide the best \textit{a priori} for our purpose (see sub-section \ref{SS:quad}). The gases included in the model are hydrogen, helium, methane ($^{12}$CH$_4$, $^{13}$CH$_4$ and CH$_3$D), acetylene (C$_2$H$_2$), ethane (C$_2$H$_6$), methylacetylene (C$_3$H$_4$), diacetylene (C$_4$H$_2$) and carbon dioxide. The different isotopes of methane were kept constant with respect to the vertical profile of $^{12}$CH$_4$. The CH$_3$D/CH$_4$ ratio used was $3.0\times10^{-4}$ as in \cite{orton2014mid2} corresponding to a D/H ratio of $4.4\times10^{-5}$. The $^{13}$C/$^{12}$C value used is the terrestrial value of $1.12\times10^{-2}$ \citep{marty2013}.

\begin{figure}[ht!]
\centering\includegraphics[width=1\linewidth]{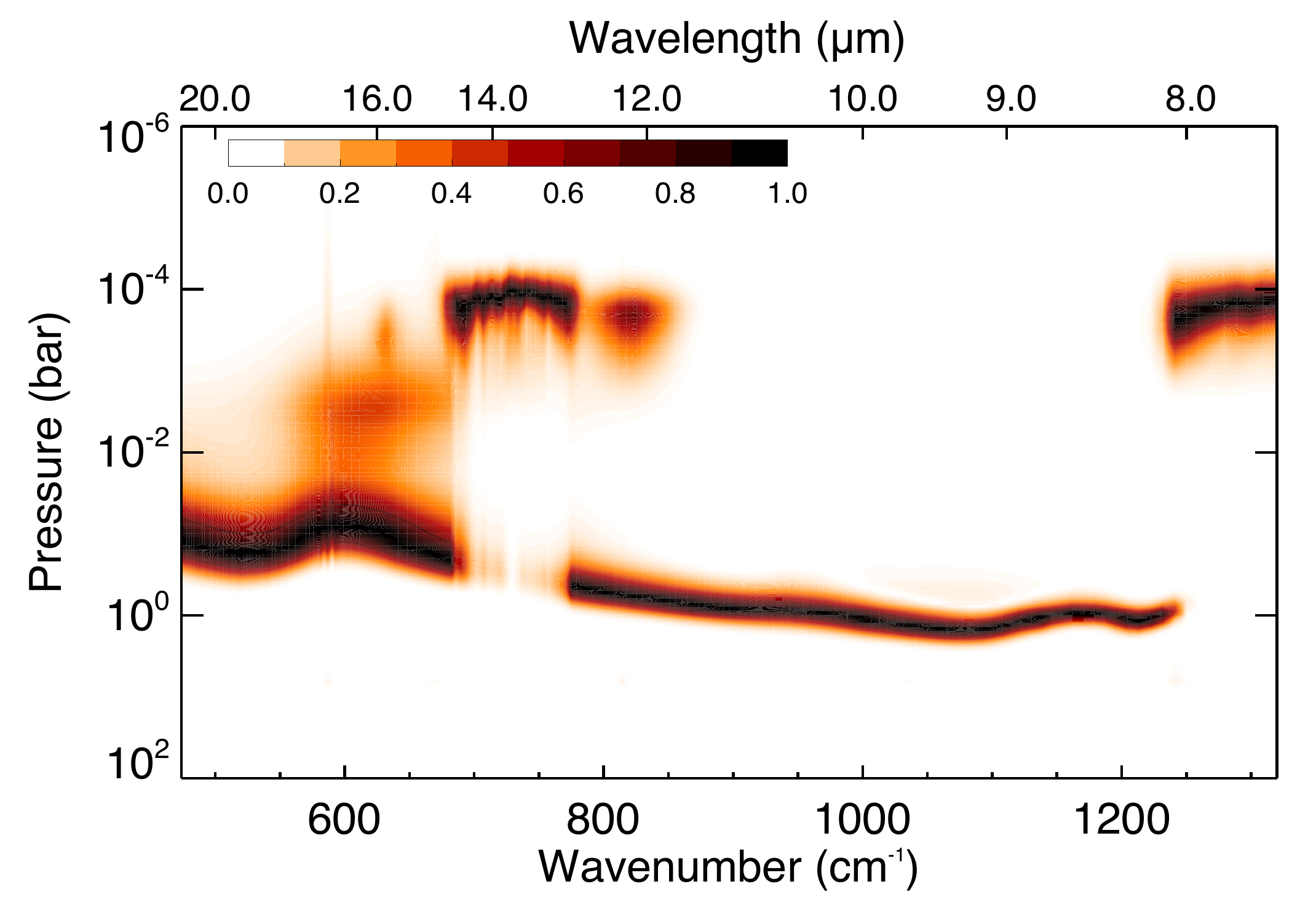}
\caption{Low-resolution contribution function contour plot showing sensitivity to different altitudes for different portions of the disc-integrated spectrum for the reference atmosphere. The top x-axis shows the corresponding wavelengths in microns. Forward modelled using NEMESIS.}
\label{fig:cf}
\end{figure}

The contribution function shows the vertical sensitivity of the Spitzer spectra, and is the product of the transmission weighting function and the Planck function (shown in Fig. \ref{fig:cf}). These are the altitudes we are most likely sensing at each low-resolution wavelength modelled by NEMESIS. At this resolution we find we cannot fit the narrow emission features of the hydrogen quadrupole without changes to the temperature prior (see sub-section \ref{SS:quad} for details). The wavelengths at which these features occur have been omitted from any subsequent low-resolution retrievals. The quadrupole lines that are in the selected high-resolution range have been, instead, covered in the high-resolution retrievals and in subsequent upper stratospheric testing. These features include the H$_2$ quadrupole S(1) line at 587 cm$^{-1}$ (17.0 $\mu$m), S(2) at 814 cm$^{-1}$ (12.3 $\mu$m), S(3) at 1034 cm$^{-1}$ (9.7 $\mu$m) and S(4) at 1246 cm$^{-1}$ (8.0 $\mu$m) \citep{orton2014mid1}. The S(0) line, present in the LH data at 28.2 $\mu$m (354 $cm^{-1}$), has also be omitted from analysis due to us choosing not to use the data beyond 19 $\mu$m for the retrievals.

The calculation of the contribution function requires the abundances of the principal absorbers in the spectral intervals chosen. The molecular abundances are derived from those as presented in \cite{orton2014mid2} (appendix A) with some small modifications made throughout the process to improve the fit. We use the nominal gas model directly from \cite{orton2014mid1,orton2014mid2} as our \textit{prior}, then allow NEMESIS to vary temperature, acetylene and ethane to fit the low-resolution spectrum. These small deviations from the \cite{orton2014mid1,orton2014mid2} model are discussed in subsequent sections.

The sources of spectroscopic linedata used in the NEMESIS model are listed in \cite{fletcher2016} Table 2. However, the underlying collision-induced absorption (CIA) of H$_2$-H$_2$ is taken from \cite{fletcher2018} so as to include the contributions of hydrogen dimers. We are assuming thermochemical equilibrium para-H$_2$ fraction throughout this work.

The low-signal region in the SL spectrum near 8-8.5 $\mu$m contains multiple sharp features that are not associated with real gaseous bands and therefore were not included in the model constraints. This region is sensitive to the carbon-13 isotopologue of methane and part of the monodeuterated methane band. The 5 to 7.4 $\mu$m section of the SL region was removed due to uncertainties in the available line data, excessive noise on the data, and the complex contribution of reflected sunlight at the shortest wavelengths. This wavelength range at Uranus has not been the subject of investigation in the past and will be the topic of subsequent analysis.

The errors used to constrain the spectral fit are the same as the measurement errors stated in section \ref{S:obs}. The only limitation to the use of percentage errors is that the inversion is weighted towards the low radiance data. This biases the findings by not allowing the model as much freedom for what are usually noisier regions of the spectrum. Additional forward modelling error ($0.019$ nW/cm$^{2}$/sr/$\mu$m) was added to the methane band between 7 and 9 $\mu$m to allow the fit to the especially noisy, low-radiance regions more freedom during the retrieval. The subsequent fit to the data is close enough for this method of errors to be considered adequate. The uncertainties on the \textit{priors} were set to 100\% of the mole fraction for each gas (at each pressure level), with temperature \textit{a priori} uncertainties of $\pm3$ K to prevent substantial deviation from the model of \citet{orton2014mid1}. The smoothing of the retrieved profiles is controlled by the correlation length parameter that is in terms of the logarithm of the pressure. A value of 1.5 is assumed for all profiles after being tested to approximate a scale height in vertical resolution. The typical number of iterations for the retrievals to converge on our reasonable chi-squared value is nine for the low resolution and five for the high resolution.

\subsection{K-Distributions}
NEMESIS is capable of generating synthetic spectra using line-by-line and also correlated-k approximation methods.  The previous analysis of \citet{orton2014mid1} used the former, but to compute the whole low resolution spectrum efficiently it was necessary to use the correlated-k method. This approach uses a Gaussian quadrature scheme to approximate the definite integral of the the rapidly varying absorption-coefficient function. The approximation is a smoother, more easily integrated function of the absorption coefficients that is a weighted sum of function values at specified points within the domain of integration. This is the k-distribution \citep{goodyyung1989}.

The number of quadrature points to integrate over (g-ordinates) is chosen to achieve the best trade-off between accurate sampling and computational speed \citep{irwin2008nemesis}. Tables computed at Spitzer SL resolution with 20 g-ordinates for integration did not reproduce the opacity in the very fine line-features such as the quadrupole features and in some of the hydrocarbon features. Increasing the resolution of the k-tables improved this a small amount, but increasing the number of g-ordinates to 50 has proven to be the most effective at reproducing the observed spectrum at both high and low-resolution.

The k-tables were generated from a high-resolution line-by-line spectrum with sufficient sampling to capture the narrow Doppler-broadened lines at low pressures, based on a correction identified by the erratum of \cite{roman2020}.

\subsection{Disc Averaging}
NEMESIS can compute numerous different types of spectra. We have chosen to compute the integrated spectral power of the planet with units of W/$\mu$m. To convert the spectral radiance units (W/cm$^2$/sr/$\mu$m) into these, a Uranian radius of 26000 km is assumed. This adds a 441 km limb extension of the atmosphere above the 1-bar pressure level at radius 25559 km. The atmospheric level that we have defined the zenith angle of the observation from is the 0 km (1 bar) altitude.

To simulate the disc-averaged spectrum we used the same method used in \cite{teanby2013uranus} as detailed in \cite{teanby2013disc}. Ten discrete atmospheric paths are used with finer spacing near the limb to account for rapid changes due to limb brightening and darkening. This number of points proved the best when balancing computation speeds and spectrum reproducability. NEMESIS calculates the spectrum along these ten paths and sums them up using the weighting in Table \ref{tab:donut}. The weighting is:

\begin{equation}
w_i=\frac{x_i x_{i+1} - x_{i-1} x_i}{r^2}
\end{equation}

Where $x$ is the offset from the sub-observer point and is associated with an emission angle. Equation 1 is adapted from \cite{teanby2013disc} using the same Uranian radius, $r$, that includes the 441 km limb from \cite{teanby2013uranus}.

This method was chosen over the NEMESIS variant that computes the disc-average using the exponential integral technique due to better model fits to the data. This variant was designed for use with secondary-eclipse spectra of exoplanets, but was found to be unable to reproduce the coldest limb-darkened parts of the Uranian spectrum, and proved too sensitive to the choice of lower pressure boundary that represents the planetary limb \citep{orton2014mid1}. The method chosen is slower but much more effective in allowing more reliable coverage of the limb, taking into account the limb brightening and darkening effects just as the 10-stream trapezoidal quadrature method in emission angle cosine used by \cite{orton2014mid1}.

\begin{table*}[ht!]
\centering
\small
\begin{tabular}{l l l l l}
\hline
i & x (km) & Tangent altitude (km) & Emission angle (\degree) & $w_i$\\
\hline
1 & 0 & - & 0.0 & 0.0 \\
2 & 5000 & - & 11.28 & 0.08876 \\
3 & 12000 & - & 28.00 & 0.23077 \\
4 & 18000 & - & 44.77 & 0.26627 \\
5 & 22000 & - & 59.40 & 0.19527 \\
6 & 24000 & - & 69.88 & 0.10651 \\
7 & 25000 & - & 77.99 & 0.05766 \\
8 & 25559 & 0 & 90 & 0.02269 \\
9 & 25600 & 41 & 90 & 0.00913 \\
10 & 25800 & 241 & 90 & 0.01527 \\
11 & 26000 & 441 & 90 & 0.00769 \\
\hline
\end{tabular}
\caption{Field-of-view averaging points used to produce the synthetic reference spectra, adapted from \cite{teanby2013disc} where $w_i$ is the weight calculated using equation 1.}
\label{tab:donut}
\end{table*}

\subsection{Eddy Diffusion}
\label{SS:eddy}

Vertical transport on Uranus is assumed to operate via eddy and molecular diffusion. Eddy diffusion is the process of the atmosphere mixing due to turbulence and the eddy diffusion coefficient profile is a major free parameter in chemical models. We use the assumption that the coefficient is the same at every altitude for Uranus as \cite{moses2018}. There is a significant effect from the vertical eddy diffusion coefficient ($K_{zz}$) on the distribution of gases in the Uranian stratosphere \citep{moses2018}. A nominal value of $K_{zz}=2430$ cm$^2$s$^{-1}$ with a methane tropopause mole fraction ($f_{CH4}$) of $1.6\times 10^{-5}$ is stated in \cite{orton2014mid2}, but they could not independently constrain these values. Different $K_{zz}-f_{CH4}$ pairs were found to fit the data with higher $K_{zz}$ values needing less methane and vice-versa. The two extremes of the $K_{zz}$ values suggested by \cite{orton2014mid2} Fig. 4 are 1020 cm$^2$s$^{-1}$ and 3270 cm$^2$s$^{-1}$. These values, and the nominal $K_{zz}$, were tested for the global retrieval. The different $K_{zz}$ values influence the shape of the vertical profiles of the main gases. For the retrieved gases, acetylene and ethane, the retrieved profiles fit the differing \textit{priors} with no significant changes to the fit of the spectrum or the temperature. All three of these values were acceptable eddy diffusion coefficients and agrees with the findings in \cite{orton2014mid2} and is the reason we chose to use their nominal value.

\subsection{Non-Local Thermodynamic Equilibrium}

The current model assumes Local Thermodynamic Equilibrium (LTE). Evidence suggests that this is suitable for the gas giants at these wavelengths but not for the extreme cold of the ice giants \citep{orton2014mid2}. At low pressures LTE breaks down and there are departures from the usual Boltzmann distribution for populations of states. This means corresponding atmospheric temperatures can differ significantly from the local kinetic temperature assumed \citep{appleby1990}. 

\cite{appleby1990} found that a temperature difference of up to 20 K is possible due to non-LTE effects at pressures of around 0.1 $\mu$bar. This is higher than the Spitzer data is sensing but significant deviations from LTE for methane at 1300 cm$^{-1}$ were found to occur starting at pressures of around 0.1 mbar. This could explain some inconsistencies in the stratospheric fitting of the retrieval.

The analysis that follows assumes LTE, which could generate a systematic offset in retrieved temperatures and abundances at low pressures if all longitudes are assumed to be uniformly affected by non-LTE processes. As the source functions for the hydrocarbons remain poorly constrained under Uranian conditions, a hybrid radiative transfer code combining LTE and non-LTE effects is planned for future work.

\section{Model Fitting and Interpretation}
\label{S:results}

\subsection{Global Retrieval}

\begin{figure}[ht!]
\centering\includegraphics[width=1\linewidth]{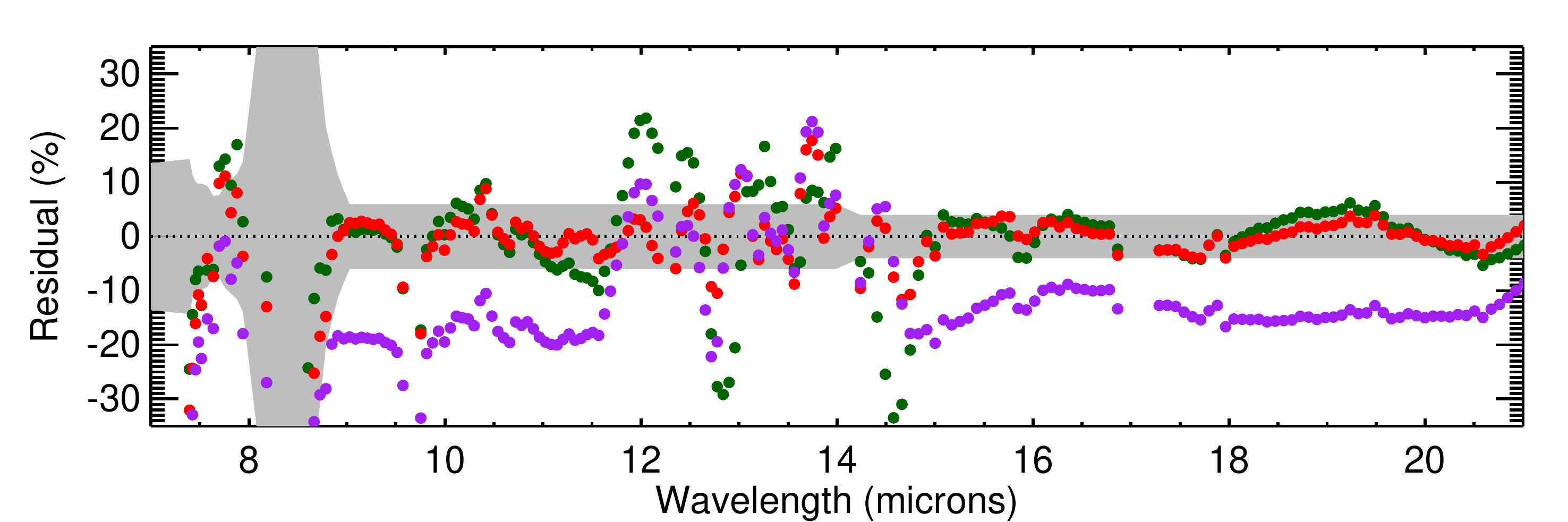}
\caption{The residuals between the low-resolution data and the retrievals for temperature change only (green circles), acetylene and ethane abundance change only (purple circles) and the temperature, acetylene and ethane retrieval used in subsequent analysis (red circles). Retrieval errors are shown as the grey shaded regions.}
\label{fig:retrsteps}
\end{figure}

\begin{figure*}[ht!]
\centering\includegraphics[width=0.8\linewidth]{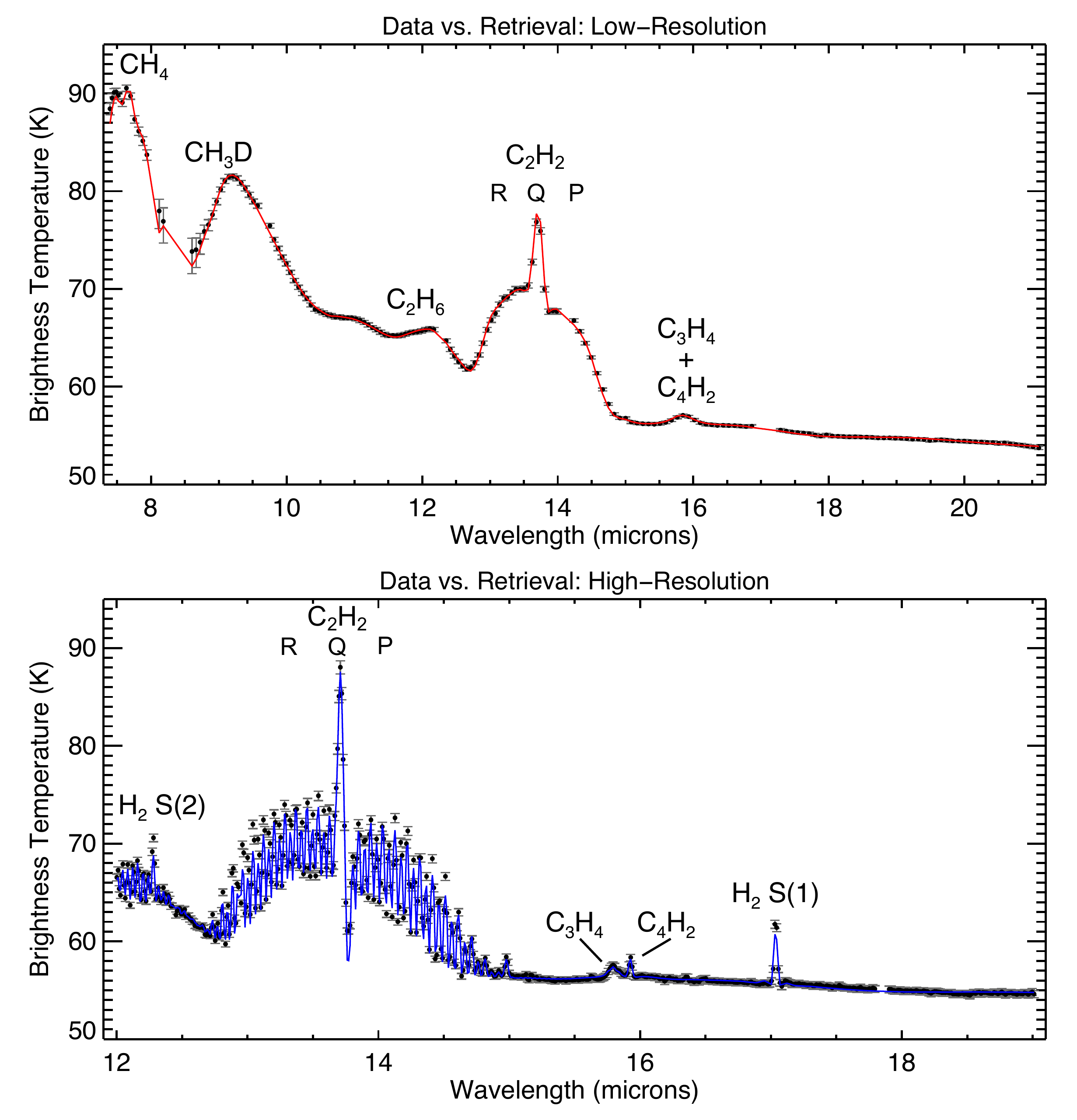}
\caption{Full spectrum retrievals for the Spitzer-IRS Uranus-2007 global averaged spectra (black dots with associated errors). The top panel shows the low-resolution (SL and LL) modules between 7.4 and 21.2 $\mu$m (red) where temperature, acetylene and ethane are allowed to vary freely. The methane region between 7.4 and 9 $\mu$m has had a fixed forward-modelling error added to account for uncertainty due to noise.  The bottom panel shows the high-resolution (SH) module between 12 and 19 $\mu$m (blue) where temperature, acetylene, ethane and diacetylene are allowed to vary freely.}
\label{fig:retr1}
\end{figure*}

The retrieval of the globally-averaged low-resolution data allows the full profiles of temperature, acetylene and ethane to vary. We retrieved from spectra in the range between 7.4 $\mu$m and 21.1 $\mu$m (472 - 1350 cm$^{-1}$) with the hydrogen quadrupole features and the noisy region around 8 $\mu$m omitted as stated in section \ref{S:method}. We could not achieve an adequate fit when allowing the acetylene and ethane abundances or temperature profile to vary in isolation, and so we needed to allow all three parameters to depart from the \textit{prior} profiles of \cite{orton2014mid2} (Fig. \ref{fig:retrsteps}). The best-fit retrieval, that allows variation from both sets of parameters, can be seen in the top panel of Figure \ref{fig:retr1} with a chi-squared value of 0.77. 

\begin{figure}[ht!]
\centering\includegraphics[width=1\linewidth]{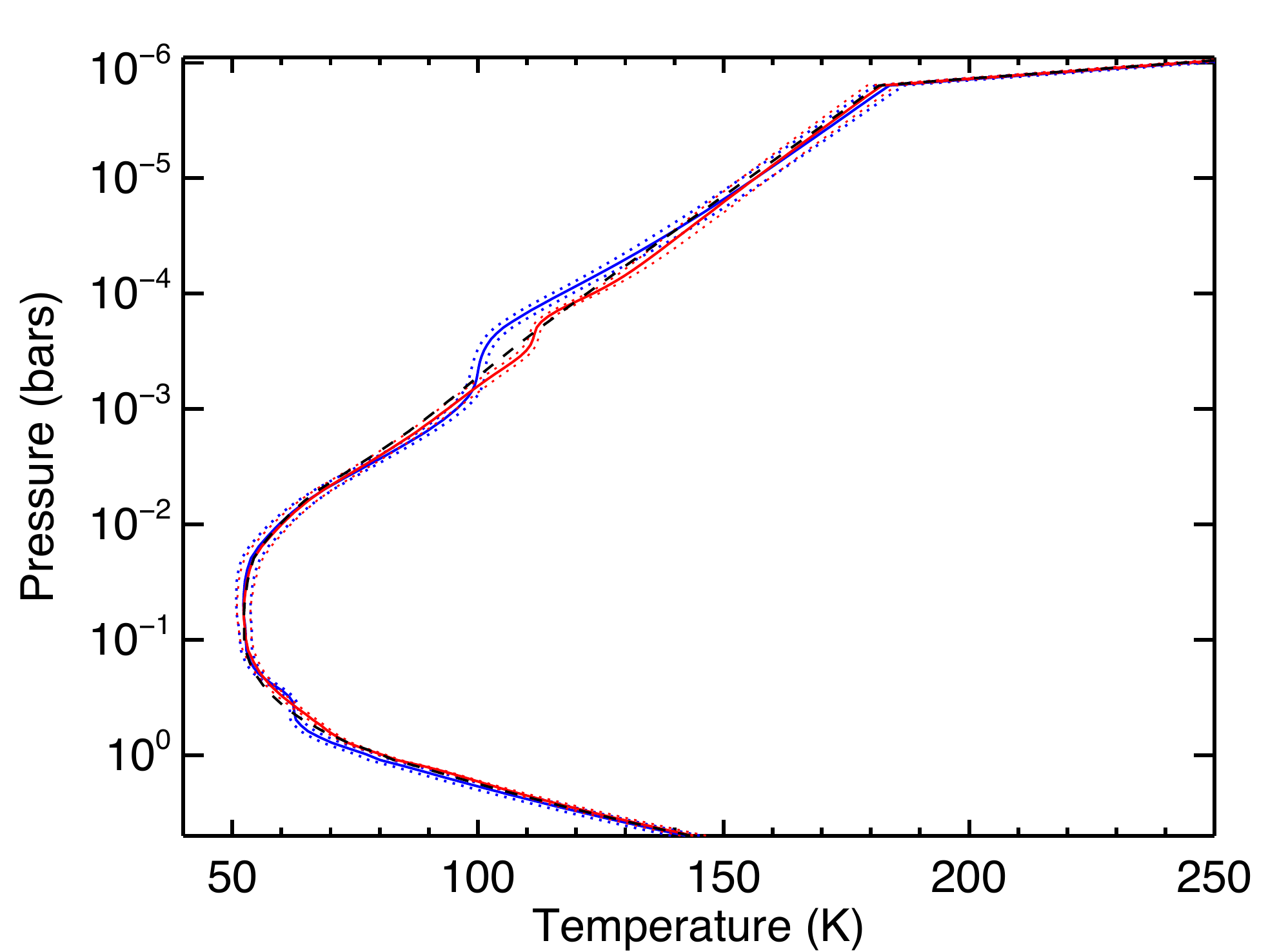}
\caption{Retrieved vertical temperature profile for the global average of the low-resolution data (red solid line) with retrieval errors (red dotted line) compared to the \textit{prior} (black dashed line). The high-resolution retrieved profile is also displayed (blue solid line).}
\label{fig:retr2}
\end{figure}

The retrieved temperature profile (Fig. \ref{fig:retr2}) is consistent with the \textit{a priori} from \cite{orton2014mid1} at most altitudes. A 3 K and 2 K increase occurs at 0.4 mbar and 1.75 bar respectively. A 2 K cooling occurs at 0.1 mbar with other deviations within the retrieved errors. We also tested alternative temperature profiles. An isotherm at $p<0.1$ mbar was proposed by \cite{orton2014mid1} based on the findings from \cite{lindal1987}. We tested similar isotherms at different temperatures and found the best fit at 110 K, but the retrieval showed the need for 7 K heating just above the 0.1 mbar level, thus diverging from the isothermal shape. We discuss the impact of this profile (and other alternatives) to the upper-stratosphere and the hydrogen quadrupole in sub-section \ref{SS:quad}.

\begin{figure}[ht!]
\centering\includegraphics[width=1\linewidth]{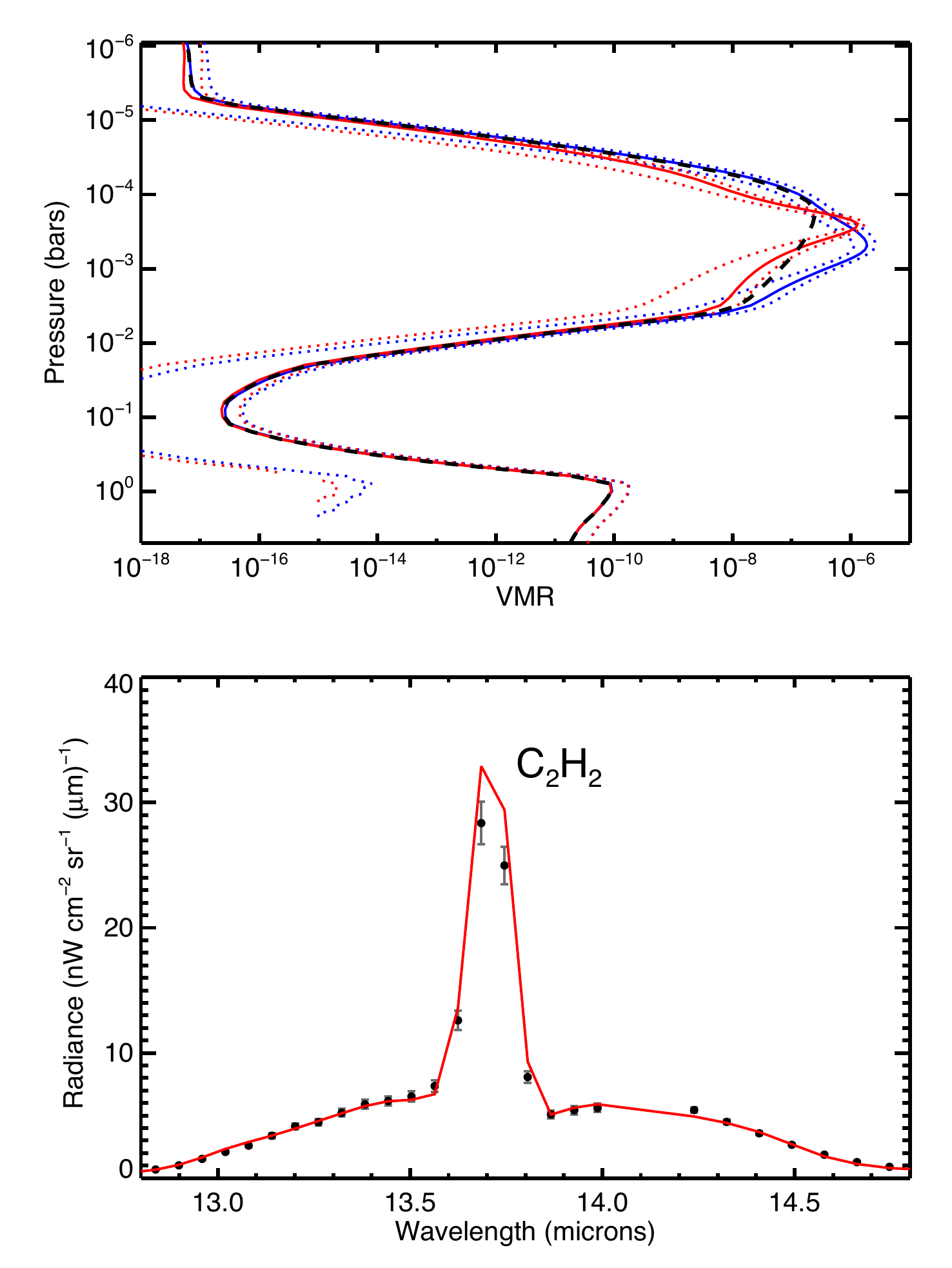}
\caption{Acetylene vertical profile retrieval and spectral fit to the global average of the data. Low-resolution retrieval result shown as red solid line in both panels. The top panel shows the volume mixing ratio (VMR) vertical profile where the \textit{prior} is shown by the black dashed line and the retrieval uncertainties shown as red dotted lines. The high-resolution retrieval result is shown for comparison (blue solid line). The bottom panel shows the spectral fit to data (black dots) with error bars for the relevant wavelength range for acetylene. }
\label{fig:zoomacet}
\end{figure}

%acetylene & ethane
The P and R branches of acetylene fit well in the low resolution but the fit to the Q-branch has a higher radiance than the data, at the upper limits of the errors (Figure \ref{fig:zoomacet}). The retrieved profile has a significant increase in abundance compared to the \textit{prior}, peaking at 0.3 mbar. This peak is slightly deeper than the \textit{a priori} peak. Above and below the new peak there is depletion relative to the \textit{prior}. The fit to the acetylene range of the spectra is within errors apart from in the Q-branch, where the model overshoots the data. This is likely due to the lack of resolution for the sharp peak and the potential for saturation at this most radiant point in the data. 

We tested the correlation-length parameter inside NEMESIS and a standard value of 1.5 was chosen. This parameter controls the smoothing of the profile fits. Higher values, that allowed more smoothing, had no significant effect on any of the profiles, including the overall chi-squared. If we force heavy smoothing of the acetylene profile, this results in an unusual vertical temperature structure. The multivariate problem is complex, and in this case we attempt to achieve a reasonable balance between the smoothness of both profiles. When we tested the acetylene, ethane, and methane priors resulting from the extremes of the vertical eddy diffusion coefficient in \cite{orton2014mid2}, we found that both solutions converged on the same abundance profile as the nominal model. Furthermore, scaling the acetylene prior to as little as 0.1 times and as much as 10 times the nominal acetylene abundance, we also found a convergence to the same nominal profile, lending confidence that our retrieval is driven by the data rather than sensitive to the prior, at least over the 0.1-1.0 mbar range.

\begin{figure}[ht!]
\centering\includegraphics[width=1\linewidth]{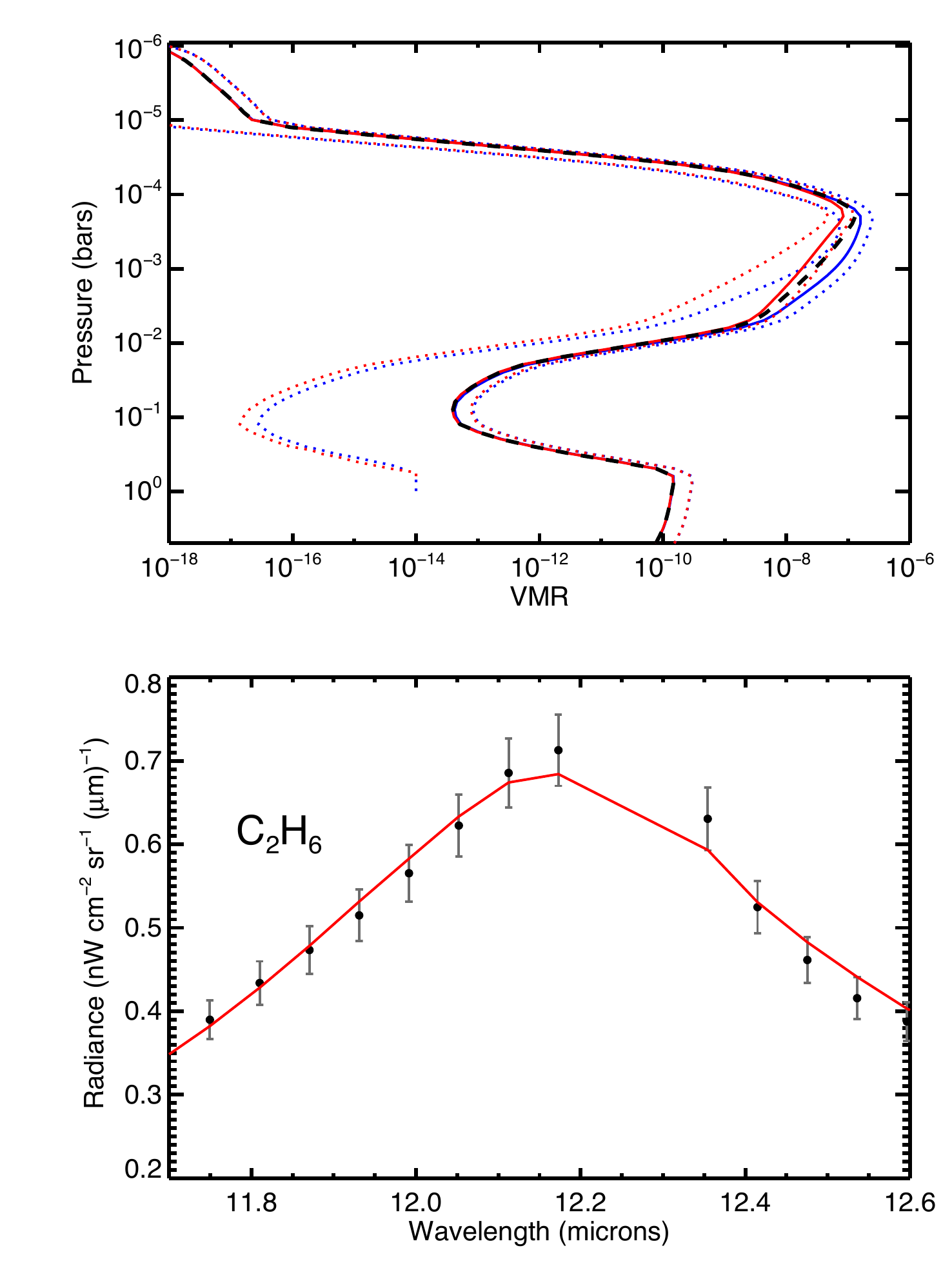}
\caption{Same as Figure \ref{fig:zoomacet} but for ethane.}
\label{fig:zoomethane}
\end{figure}

The ethane profile shows a depletion in the maximum stratospheric abundance at around 0.2 mbar with a fit within errors across the entire wavelength range (Fig. \ref{fig:zoomethane}). 

\begin{figure}[ht!]
\centering\includegraphics[width=1\linewidth]{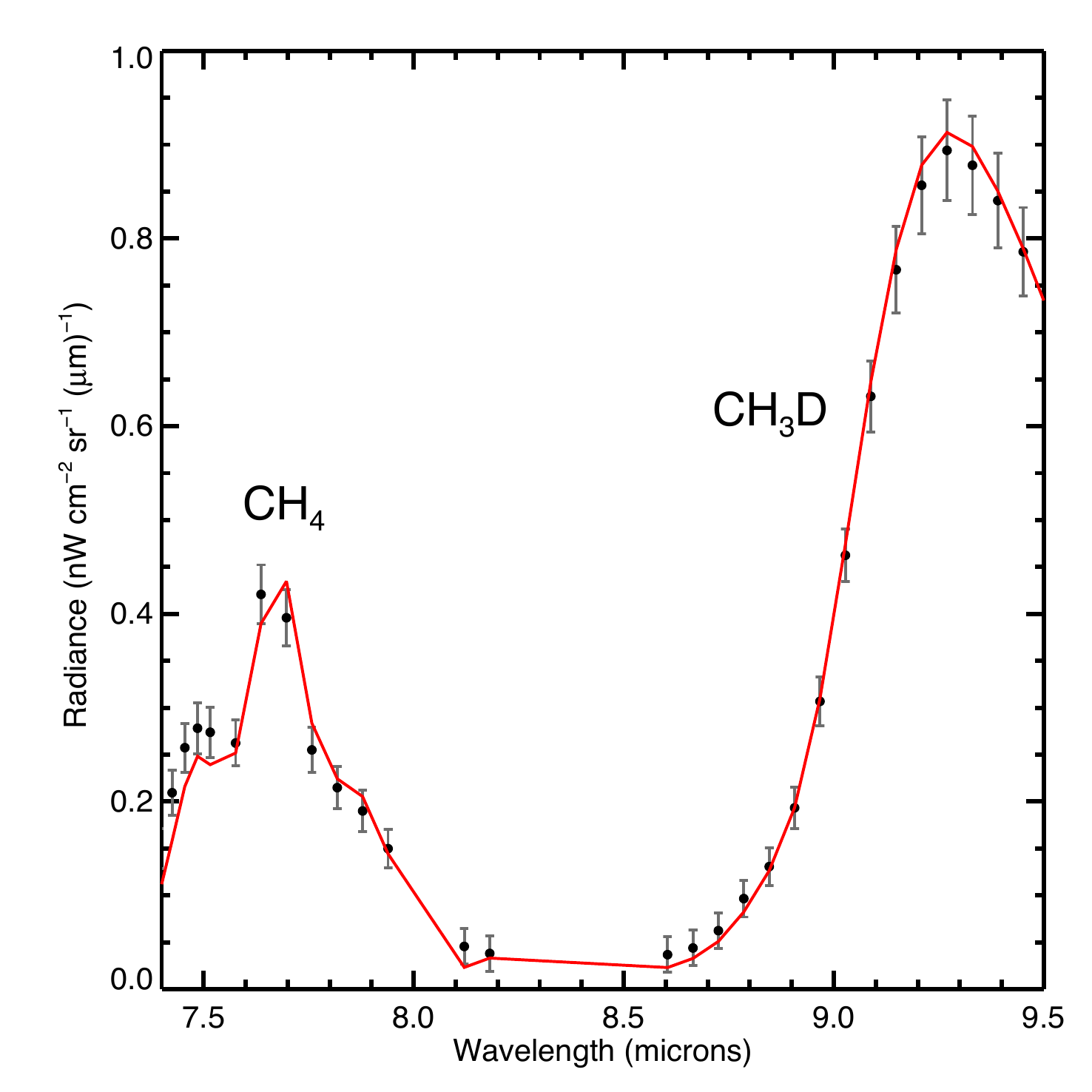}
\caption{Methane retrieval fit to the low-resolution data (black dots with associated errors) for the relevant wavelength range for both methane variants labelled.}
\label{fig:zoommeth}
\end{figure}

%methane
The spectral fit shows a difficulty in fitting the 7.4 - 7.6 $\mu$m region (Fig. \ref{fig:zoommeth}) that was also seen in \cite{orton2014mid2}. They suggested that this is caused by other emission features being present and the model not containing the necessary line data. Our fit to this region is similar in shape and so we can assume that the same problem persists in our model. The forward modelling errors added to the spectrum across the methane range were necessary to allow the temperature profile to fit realistically. This region is one of the dimmest in the entire spectrum (refer back to Figure \ref{fig:fullspec} panel B) and so the larger uncertainty is expected.

%etc
All deviations from the \textit{priors} greater than their associated errors occur in the same region of the stratosphere between 1 mbar and 0.01 mbar. The temperature and hydrocarbon fits are degenerate so we would expect any deviations from one profile of \citet{orton2014mid1} to also affect the other.

\subsection{Longitude Retrievals}

\begin{figure}[ht!]
\centering\includegraphics[width=1\linewidth]{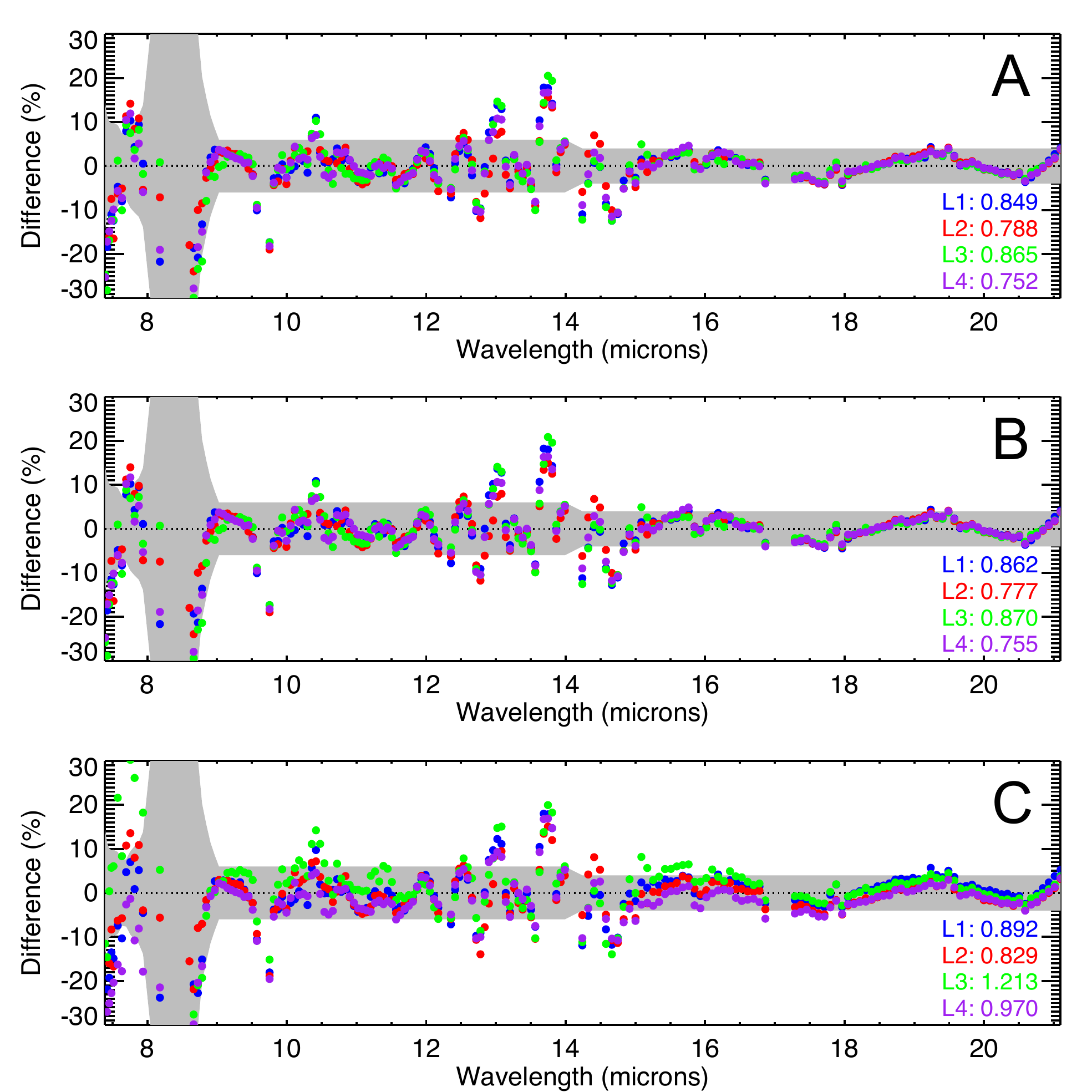}
\caption{Percentage residuals between the low-resolution spectra and the retrieved fits for each individual longitude. The three panels show three different assumptions made during the modelling process. Panel A shows the results when both temperatures and gases are allowed to vary, panel B shows temperature change only and panel C shows gas change only. Longitude 1, 2, 3 and 4 are shown in blue, red, green and purple respectively. Global-average retrieval errors are shown by the grey shaded region. The chi-squared value for each longitude retrieval is shown in the bottom right corner of each panel.}
\label{fig:hypo}
\end{figure}

The previous section demonstrated that NEMESIS is capable of fitting the globally-averaged Spitzer spectrum at low resolution with vertical profiles that are slightly modified from those of \citet{orton2014mid1,orton2014mid2}.  We now investigate the four individual longitudes, fitting each low-resolution spectrum separately. Panel A in Figure \ref{fig:hypo} shows the difference between the retrieved spectra and the data if we allow temperature, acetylene and ethane to vary. When all parameters are free, they vary together due to the degeneracy between temperature and gas abundance. To test whether the change could come from changes to gases or temperatures alone we set up two additional experiments. These are the same we tested for the global average (Fig. \ref{fig:retrsteps}) but this time the chosen parameter is fixed to the global average retrieval result and run with the other parameter as the only retrieved variable. Panel B shows the residuals when temperature is the only parameter changing. Panel C shows the residuals where only acetylene and ethane are allowed to vary.

Hypothesis A and B (Fig. \ref{fig:hypo}) have residuals and chi-squared values that are almost identical to each other. Panel C, however, shows that the model is less able to fit the data within errors when only gases are varied, and has discrepancies in the fits between longitudes. The chi-squared values are higher for longitudes 3 and 4 that exhibit the largest variation from the average. 

Temperature variations are all that is required to fit the spectra to the same accuracy as when we change all of the variables. There is no known reason for chemical abundances to change with longitude, as photochemical time scales are extremely long, except perhaps if there are localized regions of vertical mixing. The small temperature perturbations retrieved here will have no effect on the chemical kinetics rates. We can therefore assume that temperature is the only parameter changing to cause the longitudinal variations observed in the emissions of the photochemical constituents of the stratosphere seen in section \ref{S:var}. 

\begin{figure}[ht!]
\centering\includegraphics[width=1\linewidth]{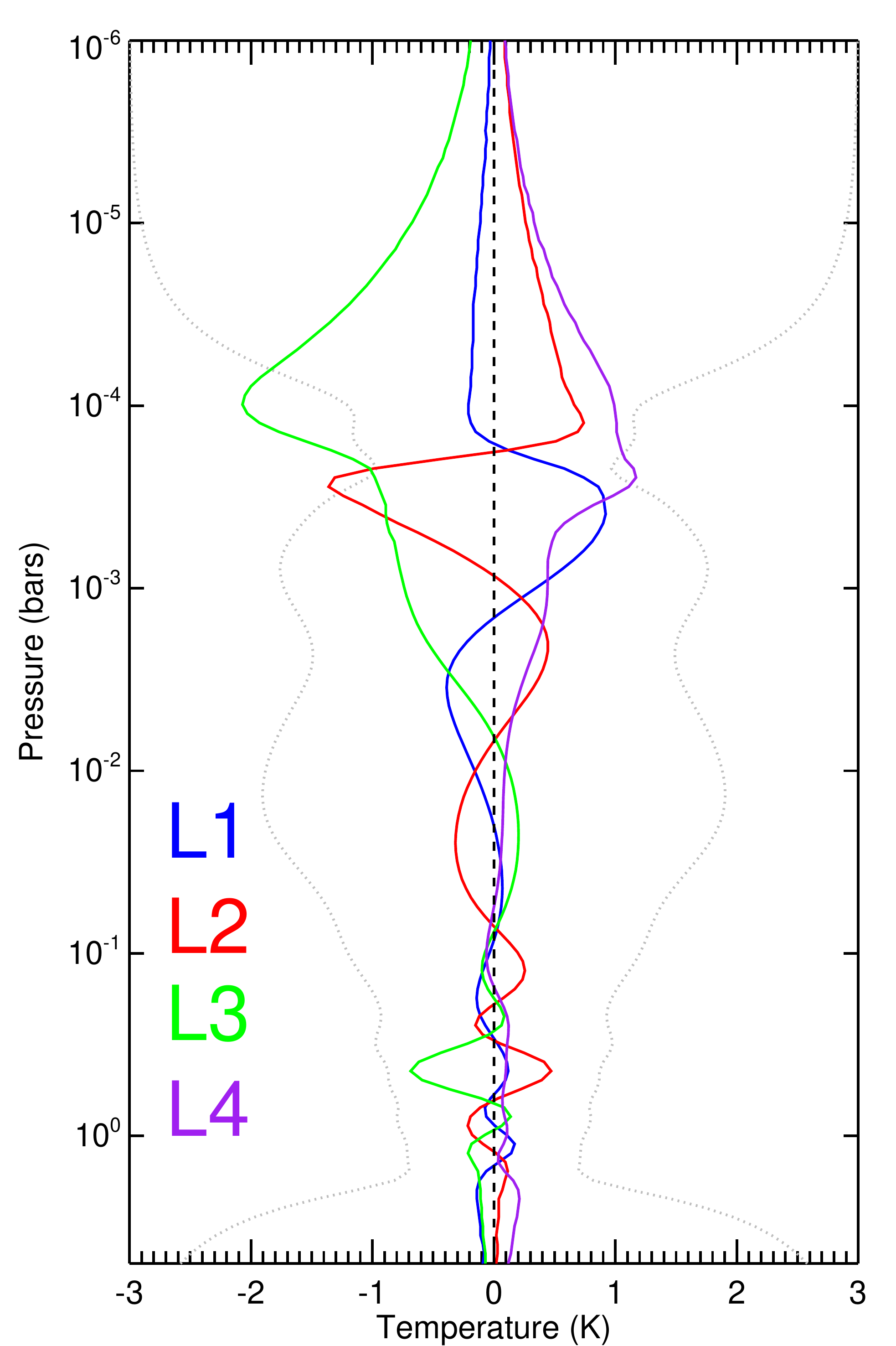}
\caption{Temperature differences from the global mean at each pressure for each longitude retrieval, assuming the temperature is the only variable that is changing (corresponding to hypothesis B in Figure \ref{fig:hypo}). Retrieved errors of the global average temperature are shown as grey dotted lines.}
\label{fig:longtemp}
\end{figure}

If we assume that the longitudinal variability is caused solely by changes in temperature, then the temperatures required to cause the spectral changes in the four longitude observations are reasonable. Figure \ref{fig:longtemp} shows a 3 K difference at 0.1 mbar between the most extreme longitudes. Longitude 3 is cooled by 2 K at 0.1 mbar. Longitude 4 is heated by 1.2 K just below the same altitude where longitude 2 also shows a cooling of 1.3 K. These three extremes of temperature change are the only altitudes at which the temperature difference exceeds the associated retrieval uncertainties of the global average retrieval.

\subsection{High-Resolution Retrievals}

In addition to the low-resolution dataset, the high-resolution observations show a similar longitudinal variability. We initially tried to forward-model at high resolution, using the low-resolution retrieval results, to see if we could reproduce the same variations that were observed in the high-resolution data. This was not successful and resulted in differences between the model and data. This offset is due to the shift in the longitude between observations with the SL and SH modules (Table \ref{tab:modtime}).The change in the brightness was too much for this test to be meaningful so we therefore attempted to invert the high-resolution data as a further test of the source of the variations.

When inverting the high-resolution data we encountered several challenges: (i) The data had to be scaled and pivoted to the low-resolution during calibration; (ii) the wavelength ranges do not contain the necessary information to constrain the variation of methane; (iii) the resolution is estimated at $R\sim600$ but is found to vary considerably as a function of wavelength \citep{orton2014mid1}. Nevertheless, here we attempt to make use of the high-resolution data to study the longitudinal variation.

We attempt to fit the globally-averaged high-resolution data to identify how they might differ from the low-resolution results. The bottom panel of Figure \ref{fig:retr1} shows our best fit to the full high-resolution spectrum. The differences between the low-resolution (red lines) and the high-resolution (blue lines) retrieval results can be seen in the temperature, ethane and acetylene profiles (Figures \ref{fig:retr2}, \ref{fig:zoomethane} and \ref{fig:zoomacet} respectively). The temperature profile shows cooling of around 7 K compared to the \textit{prior} at around 0.1 mbar. For acetylene, the deviation from the \textit{prior} is similar in magnitude to the low-resolution result but peaks slightly deeper (0.3 mbar). The cooling and the increase in acetylene in this same altitude range indicates a level of degeneracy between these two parameters that was also identified in the low resolution inversions. For ethane, instead of a depletion at stratospheric altitudes (between 0.1 and 10 mbar) there is an increase in abundance. At the 1 mbar level the ethane abundance is increased compared to the \textit{prior} by a factor of 1.3 for the high-resolution spectrum compared to a depletion by a factor of 0.3 for the low-resolution spectrum. These differences in temperature, acetylene and ethane could be real and related to the $\sim20\degree$ change in the longitude between modules (Table \ref{tab:modtime}), but could just as likely be due to the several modelling challenges we listed above. All of the retrieved profiles peak slightly deeper than in the low resolution results and could be due to differing vertical resolution causing differences in the high-resolution contribution functions.

\begin{figure}[ht!]
\centering\includegraphics[width=1\linewidth]{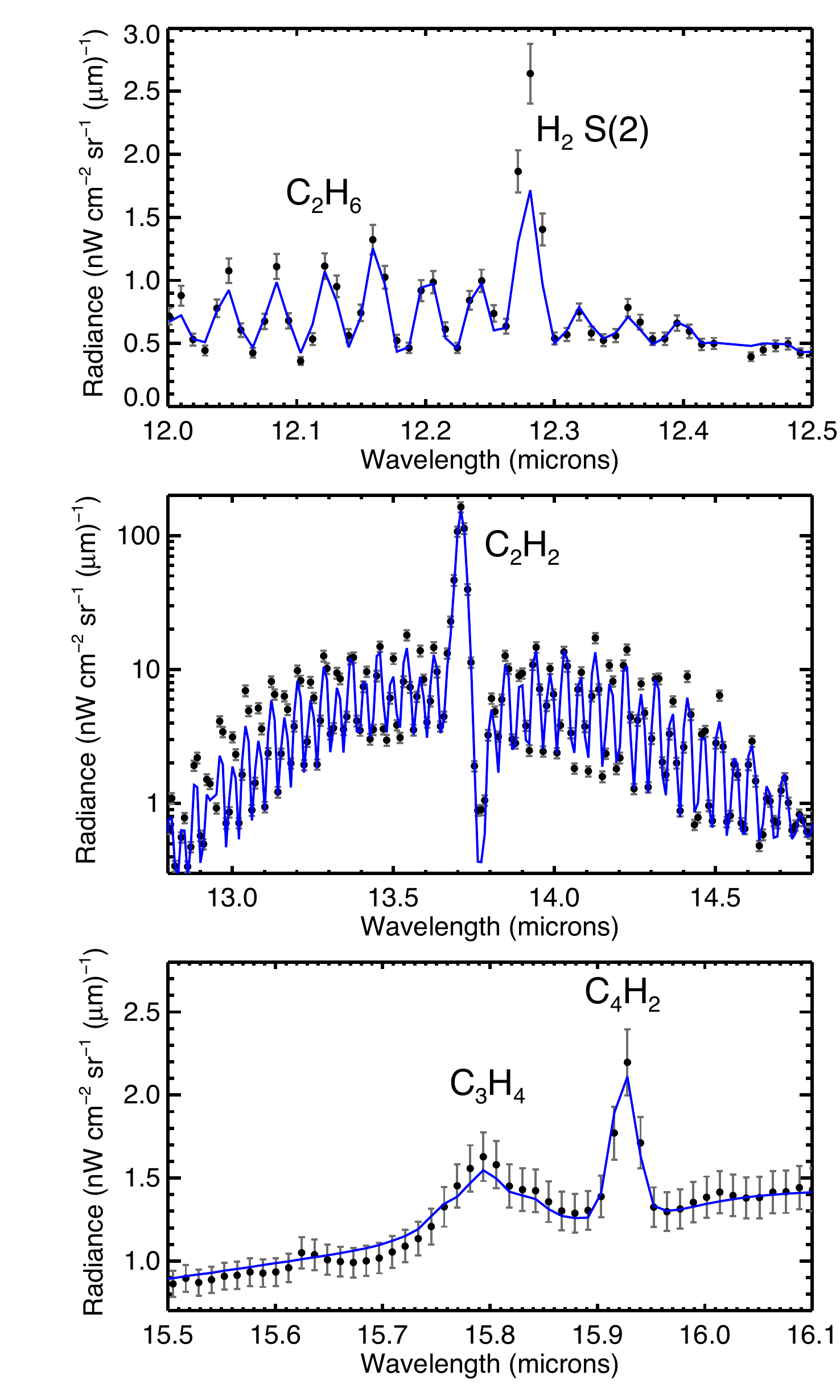}
\caption{The spectral fit of the high-resolution retrieval (blue solid lines) to the data (black dots) with 9\% error bars. Top panel: Ethane region with embedded H$_{2}$ S(2) feature. The poor fit of the model to this feature is discussed in sub-section \ref{SS:quad}. Middle panel: Acetylene P, Q and R band regions. Bottom panel: Methylacetylene and diacetylene region with unknown feature visible at 15.6 $\mu$m.}
\label{fig:zoomhigh}
\end{figure}

Fits to the spectra are shown in some important regions in Figure \ref{fig:zoomhigh}. Ethane shows a good fit across the range with the embedded H$_2$ S(2) line clearly visible. This modelled line is dim compared to the data which may suggest that the fitted temperature never quite gets warm enough to explain this emission when fitting the whole spectrum. The H$_2$ S(1) is also dim and we discuss this further in sub-section \ref{SS:quad}. 

The fitted acetylene bands P and R are consistently lower than the observed brightness. This underfitting could be caused by the k-tables being sampled at a fixed spectral resolving power of 600 even though the resolution is variable across the wavelengths. However, although several different resolution k-tables were created to try to improve this fit, no real improvements were seen. We also tried to fit narrower sub-ranges of the acetylene emission, treating the P, Q and R branches independently, but had the same issue. The low-resolution acetylene fit also had some issues, though not as severe, and this was reasoned to be caused by saturation of the bright Q-band peak. The same can be reasoned for this range in high resolution.

Diacetylene and methylacetylene (propyne) both fit well. Diacetylene had to be varied in the retrieval to stop consistently underfitting. A very small increase in abundance of a factor of 1.6 between 0.2 and 1 mbar was necessary to fit this feature but was within the retrieved errors. Carbon dioxide at 667 cm$^{-1}$ (15.0 $\mu$m) also fits within errors. The unknown feature at 640 cm$^{-1}$ (15.6 $\mu$m) was observed by \cite{orton2014mid2} and can be seen in the bottom panel of Figure \ref{fig:zoomhigh}. This same feature also shows up in ISO data of Saturn in \cite{moses2000saturn}. Another, small unknown feature is also present at 624 cm$^{-1}$ (16.0 $\mu$m). Both unknown features are within the uncertainty of the data.

The full high-resolution inversions suggest slightly modified temperatures and gas profiles compared to the low-resolution data. This deviation could be caused by the longitudes being sampled up to $20\degree$ apart (Table \ref{tab:modtime}) though the calibration of the SH data is sufficiently challenging that it is understandable that more weight is to be placed on the variations revealed by the better-calibrated low-resolution results.

\subsection{Upper-Stratospheric Temperatures and the Hydrogen Quadrupole}
\label{SS:quad}

The hydrogen quadrupole features present in the data include the S(1) line at 587 cm$^{-1}$ (17.0 $\mu$m), S(2) at 814 cm$^{-1}$ (12.3 $\mu$m), S(3) at 1034 cm$^{-1}$ (9.7 $\mu$m) and S(4) at 1246 cm$^{-1}$ (8.0 $\mu$m) \citep{orton2014mid1}. S(1), S(2) and S(3) are clearly visible in the low-resolution data. The S(4) is within the noisy range of the spectrum and therefore we have omitted it. The SH module data feature the S(1) and S(2) lines.

The quadrupole lines are all consistently underfitting in both the high- and low-resolution retrievals when we restrict ourselves to the \cite{orton2014mid1} \textit{prior}. This is why we have omitted the quadrupole spectral ranges in the global and longitudinal study of the low-resolution spectra in the previous sections, after clarifying that their omission does not impact our findings for the longitudinal variations.

\begin{figure*}[ht!]
\centering\includegraphics[width=0.8\linewidth]{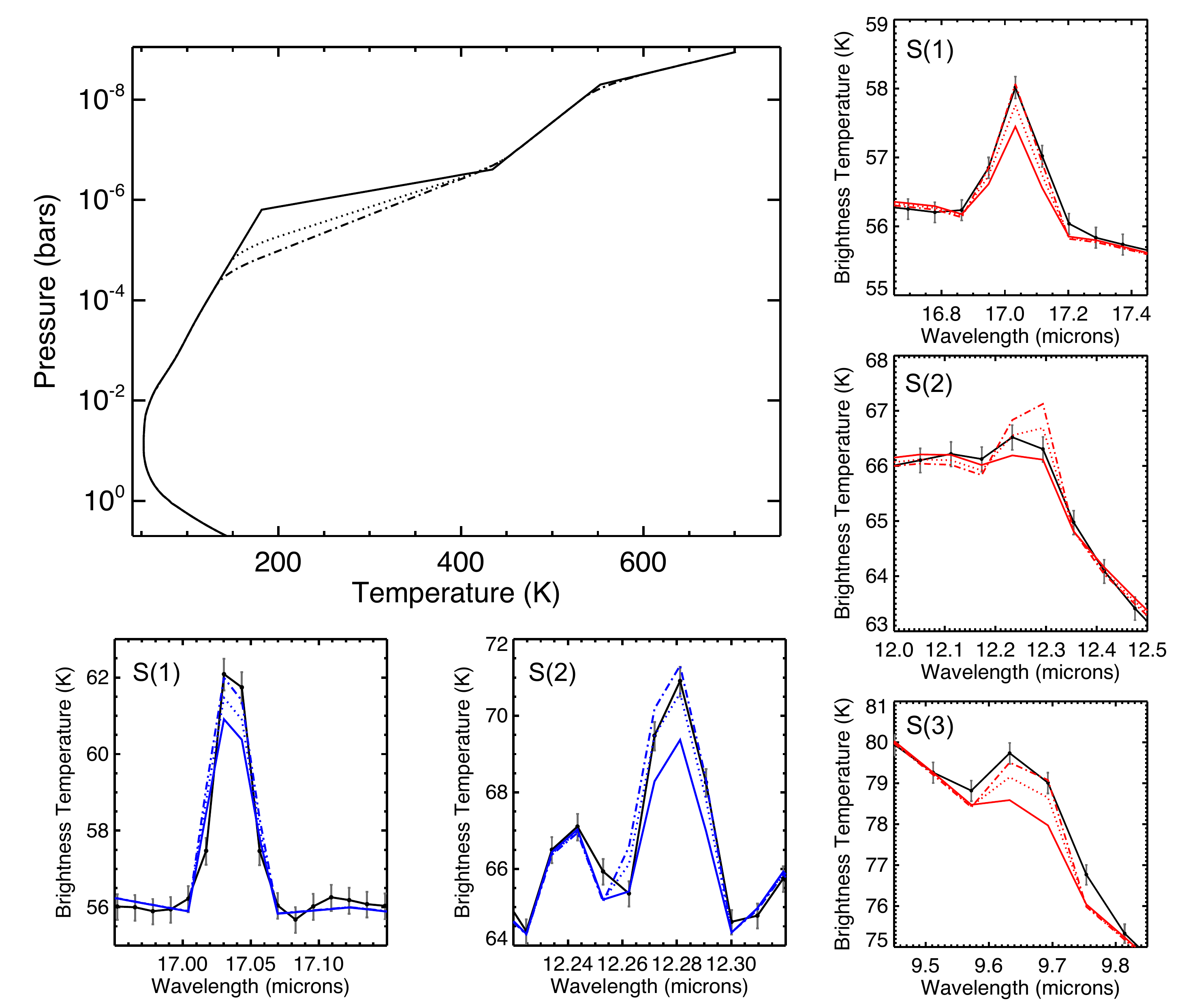}
\caption{Temperature profile \textit{a priori} that was used to test the quadrupole line fits and the upper stratosphere extended up to 1 nanobar. The nominal temperature profile with extension is shown with the solid line, the profile with a smoothed slope starting at 10 $\mu$bar is shown by the dotted line and the profile starting at 28 $\mu$bar is shown by the dashed-dot line. The quadrupole line fits during the full spectral retrieval for each temperature \textit{prior} are shown for the low resolution (red) and high resolution (blue). The line styles match the temperature profile that they represent. The quadrupole line S(1) is at 17.0 $\mu$m (587 cm$^{-1}$), S(2) at 12.3 $\mu$m (814 cm$^{-1}$) and S(3) at 9.7 $\mu$m (1034 cm$^{-1}$).}
\label{fig:tempslope}
\end{figure*}

\cite{orton2014mid1} found that to successfully model the quadrupole lines they had to extend into the upper stratosphere up to 1 nanobar. The nominal temperature profile used in the main investigation cuts off at 0.1 $\mu$bar so we have extended it in Figure \ref{fig:tempslope}. This extended temperature profile reaches 700 K in the thermosphere, consistent with Voyager UV occultation measurements \citep{herbert1987}. We increased the number of pressure levels to 200 and used an equivalent disc-averaging method that was extended up to a radius of 27000 km (a limb of 1441 km above the 1 bar level). The nominal fits to the quadrupole have the same differences to their respective spectral data as the nominal values in Figure \ref{fig:retr1}. Modelling of the quadrupole lines alone does require the upper extension to be properly accurate but when retrieving the entire spectrum it does not make a difference outside of uncertainties. As we found it made no significant difference to the full spectrum retrievals to include these upper altitudes, we tested the alternative temperature profiles in Figure \ref{fig:tempslope}.

Although we cannot fit the quadrupole lines with the \cite{orton2014mid1} \textit{prior}, we can get closer if we adjust the \textit{prior} by warming it at the upper stratospheric pressures that the quadrupole lines are sensitive to (Figure \ref{fig:tempslope}). These alternative temperature \textit{priors} allow better fits to the quadrupole features without affecting the fit elsewhere in the spectrum. The retrieved temperature fits were all within uncertainties of the nominal temperature retrieval, apart from at the altitudes that are warmed, where they are within the uncertainties of the adjusted \textit{prior}. We have displayed two of the profiles in Figure \ref{fig:tempslope}. One has a smoothed slope starting to warm beyond the nominal profile at 10 $\mu$bar and the other starting at 28 $\mu$bar. They show that whilst we can fit the S(1) and S(3) lines together,  this will then overfit the S(2) line as a consequence. We are assuming thermal equilibrium para-hydrogen fractions at the warm temperatures of the upper stratosphere, and this could be evidence that equilibrium is not the right solution for the quadrupole line fits \citep{conrath1998,fouchet2003}. The degree of disequilibrium could tell us a lot about atmospheric circulation and would be a good topic for future study.

This is why we chose to omit the quadrupole features from our low-resolution investigation and rely on the temperature \textit{prior} found from \cite{orton2014mid1}. These features only represent a small number of spectral points so the retrieval is favouring the richer data of the hydrocarbon bands, sacrificing the fit to the narrow lines. We have found that the extension into the upper stratosphere and accurate modelling of the hydrogen quadrupole do not affect the conclusions about relative changes from longitude to longitude.

As mentioned previously, we also experimented with a range of \textit{a priori} with isotherms at $p<0.1$ mbar. All \textit{priors} diverged from the isotherm shape, heating by up to 7 K just above the 0.1 mbar level. The best fit quasi-isotherm (at 110 K) was found to fit the majority of the spectrum but it makes the fit to the quadrupole lines worse. A heating of the upper stratosphere is favourable over this solution in order to fit the quadrupole lines along with the rest of the spectrum in both the high and low resolution.

\section{Discussion}
\label{S:disc}

\subsection{Comparison to Orton \textit{et al.}}

Our reanalysis of the Spitzer/IRS globally-averaged spectrum converged on similar temperature and hydrocarbon profiles as \cite{orton2014mid1}, with the exception of deviations near the 0.1-mbar level and at even lower pressures to allow us to fit the quadrupole features. This is an independent fit to the newly reduced Spitzer data with a different spectral model that largely supports the results of \cite{orton2014mid1,orton2014mid2} with some refinements. For example, Orton \textit{et al.} modelled individual features separately with a line-by-line model, whereas we have inverted the entire range simultaneously with $k$-distributions to reproduce the newly reduced, and slightly brighter, Spitzer data.

The retrieval was only successful in replicating the data with adjustments to the acetylene and ethane profiles. This suggests that there is a real deviation from the gas \textit{priors}, mainly acetylene, within this range. The increase in acetylene is required to allow NEMESIS to achieve an adequate fit across the whole spectral range. This sharp peak in abundance cannot be easily explained by photochemistry, and we cannot discount issues related to saturation at the peak of this relatively bright feature. There could also be hidden parameters contributing to this fit, such as systematic problems related to calibration of the data, and the effect of a non-uniform disc on the disc-averaging process. Non-LTE effects could also potentially explain these differences if they were factored into the retrievals. These anomalies at around 0.1 mbar are also present in the high-resolution retrievals but, with the data having such challenging reduction and calibration, they cannot be relied upon. We used this global fit as a baseline to study the relative variability that should be unaffected by any discrepancies from Orton \textit{et al.}.

\subsection{Longitudinal Temperature Changes}

Having demonstrated the longitudinal variation occurring at Uranus is real and consistent with temperature changes in the stratosphere, we discuss where these variations may have originated from. It is still possible that temperature and abundances are both changing to cause the variation, but the fact that we can simultaneously reproduce the spectral emission from several molecular species by adjusting the temperature profile alone, suggests that a simple temperature perturbation may be responsible for the observed differences between longitudes.

Temperature change in the stratosphere of giant planets has been known to originate from intense storm activity that causes upwelling of cold material from the troposphere into the lower stratosphere. Voyager 2's IRIS instrument observed an intense and localised cold region in the lower stratosphere at Neptune \citep{conrath1989neptune} that was thought to be associated with the Great Dark Spot. In 2006, a possible dark spot at Uranus was observed using HST \citep{hammel2009}. If a dark spot or other large vortex was discovered and persisted through to December 2007, it is possible that this could cause a tropospheric temperature anomaly like that observed at Neptune, but the influence of tropospheric vortices on stratospheric temperatures is uncertain. Variability seems to occur in 2005, though it is weaker and seemingly at higher pressures than in 2007. This either shows that this type of anomaly occurs at Uranus frequently, or that this is the same feature (this possible dark spot) that has persisted, evolved and strengthened through to 2007. The 2004 data is so noisy that no conclusions can be drawn about whether the variation was present at that time. More frequent and repeated observations in the mid-infrared are necessary to make future insights.

If the variation is caused by a dark spot then this would create a suspected stratospheric cold anomaly would exist at Longitude 2 and 3. We can assume that any localised cold region would be located between $0\degree$ and $180\degree$. Conversely, we cannot discount the possibility that a hot region was located in the opposite hemisphere, much like the Saturn stratospheric beacon \citep{Fletcher2012,moses2015saturnbeacon,cavalie2015saturnbeacon}.

An alternative origin for the thermal variation could be a stratospheric wave with wavenumber 1. Waves are common in the stratospheres of Jupiter \citep{fisher2016,fletcher2016}, Saturn \citep{achterbergflasar1996} and Neptune \citep{sinclair2020spatial}. It was postulated that the Great Dark Spot at Neptune was driving a stratospheric wave that then caused the cold temperature anomaly observed by Voyager 2 \citep{conrath1989neptune}.

We can not be sure if this change is isolated to one location or is a global phenomenon. JWST will provide us with the spatial resolution at mid-infrared wavelengths that is required to give us insight into what is happening. We can, however, use spatially-resolved images from ground-based instruments in the near- and mid-infrared to give us clues to this stratospheric anomaly.

\subsection{Comparison to VLT Mid-IR images Post-Equinox}

Mid-infrared images from VLT-VISIR at 13 $\mu$m \citep{roman2020} revealed warm mid-latitude bands of acetylene emission in 2009, two years after the Spitzer observations. Taken over two nights, these series of images appear to show hints of zonal variation with marginally greater emission at similar longitudes. Although the signal-to-noise ratio in these data was too low to offer conclusive evidence of longitudinal variation, similar future imaging data at higher signal-to-noise from the ground and JWST should be able to help determine whether discrete features or larger planetary waves are responsible for the variation seen in the Spitzer data.

These images, though taken years after the Spitzer observations, provide us with a clue to the possible spatial extent of the longitudinal variation observed with Spitzer. The observed features were at the edge of detectability in the ground-based 13 $\mu$m images, but appeared to be limited to the brighter mid-latitude band, and extended over half the visible longitudes. This shows it is feasible that the variation is localised on a small portion of the disc such that the 3-K contrasts we measure in the disc-averages could be a lower limit on the actual temperature perturbations in Uranus’ stratosphere.  However, any conclusions would be speculative at best until the Uranian stratosphere is observed with the spatial resolution and sensitivity of JWST.

This data suggests that tropospheric meteorology could be influencing stratospheric temperatures in subtle ways that either persist over long timescales or have been observed coincidentally two years apart. We also have tentative evidence of a connection between direct upwelling in the troposphere and the stratospheric temperatures during the time of the Spitzer observations.

\subsection{Comparison to Keck NIR images at Equinox}

\begin{figure}[ht!]
\centering\includegraphics[width=1\linewidth]{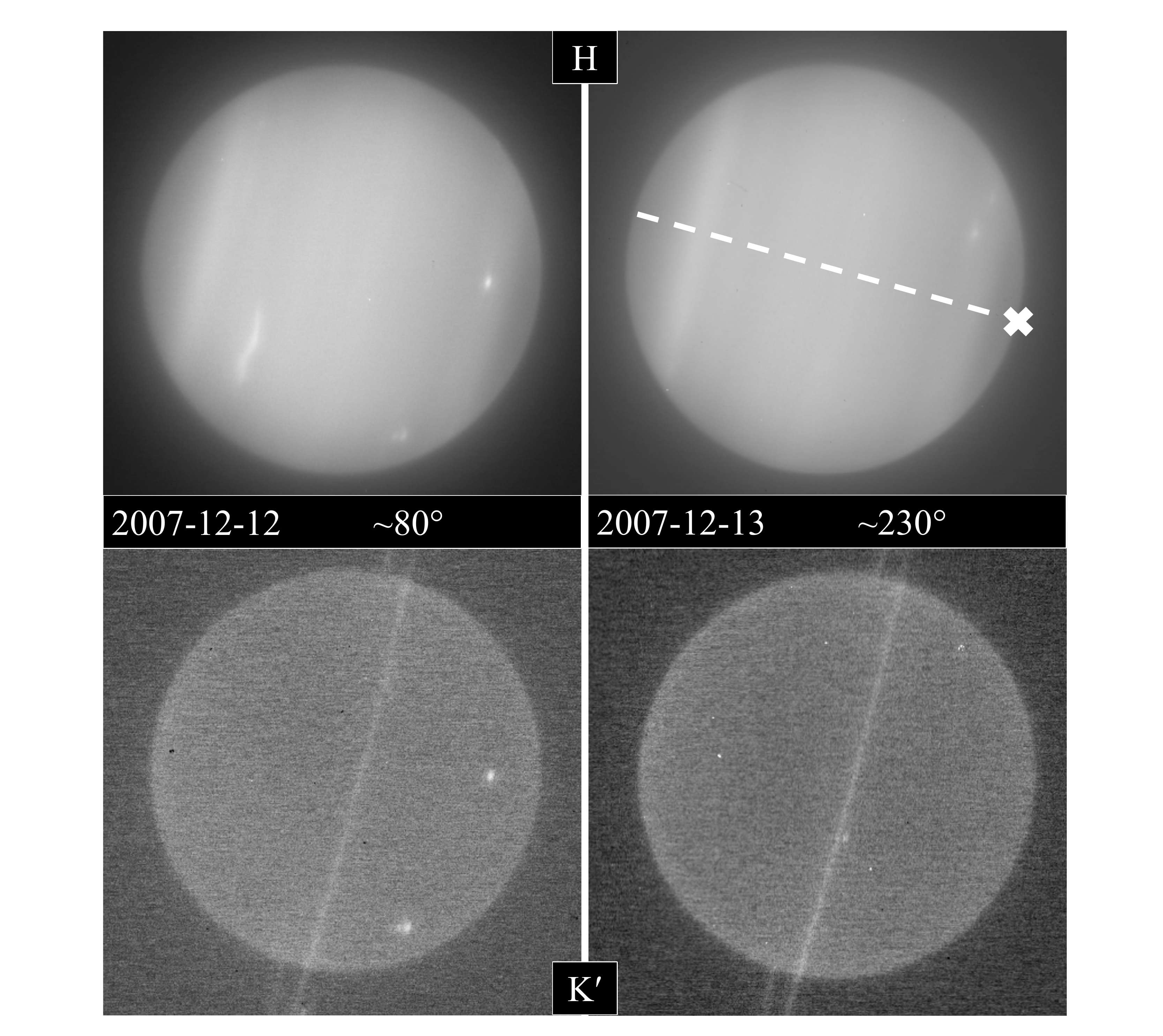}
\caption{Images of Uranus at 1.6 $\mu$m (H band, top) and 2.1 $\mu$m (K$'$ band, bottom) taken with Keck II NIRC2 instrument on two dates in December 2007. Approximate central meridian longitude of each image is displayed in degrees east \citep{depater2011,depater2013}. Orientation of the planet indicated by an X at the northern pole and a dashed line along the north-south meridian.}
\label{fig:keck}
\end{figure}

The near-infrared images in Figure \ref{fig:keck} were taken with NIRC2 coupled to the adaptive optics system on the Keck II telescope (Table \ref{tab:obssum}). They were acquired one week after the equinox and 4 days before the Spitzer data on 2007-12-12 and 2007-12-13. The H-band images sense down to several bars and are dominated by scattering at 1-3 bars \citep{sromovsky2009,depater2011}. Clouds at pressures less than 1 bar are visible in the K$'$ band \citep{depater2013}. 

The two observed dates show approximately opposite hemispheres with the 2007-12-12 observation revealing a more meteorologically active hemisphere than the 2007-12-13 observation. The activity in both spectral bands show that a few discrete cloud features exist at pressures well less than 1 bar. These clouds show regions of condensation located high above the main cloud layers and likely indicate local perturbations in the temperatures or dynamics (from below). We therefore speculate that they could also influence the stratosphere, either by direct advection of mass, or by generating waves that propagate vertically \citep[such as during Saturn's 2010-2011 storm,][]{Fletcher2012}.

The central longitudes roughly correspond to the longitudes in the Spitzer observations (Fig. \ref{fig:hemex}). We calculated the locations of the features using drift speeds at their respective latitudes assuming constant wind velocities \citep{sromovsky2009}. The minimum in the stratospheric emission and temperature at longitude 3 coincides with the hemisphere where the troposphere exhibits more meteorological activity, such that upwelling and adiabatic expansion might explain the cooler stratospheric temperatures.

However, the individual features are small compared to the overall size of the disc, and no such large temperature perturbations have been observed in upper tropospheric temperatures of Uranus to date \citep{orton2015, roman2020}.  Thus direct convective overshooting into the stratosphere seems unlikely but it is possible that one of these bright features is associated with a much larger and invisible dark spot like the one supposed in \cite{hammel2009}. The dark spot has been linked to the intense and long-lasting feature named the ``Bright Northern Complex'' (BNC) that was visible from 2005 \citep{sromovsky2007} through to 2007 \citep{sromovsky2009}. The BNC was located at around 30\degree North in August of 2007 and it is likely that it is one of the two northern-hemisphere features seen in the left panels of Figure \ref{fig:keck}. These clouds are located between 200 and 600 mbar \citep{hammel2009,roman2018} and provide circumstantial evidence that long-lived tropospheric disturbances may be related to the longitudinal temperature variations we infer in the stratosphere.

The large southern-hemisphere feature in the Figure \ref{fig:keck} H-band image is referred to as the ``Berg'' \citep{hammel2001,depater2002,sromovsky2009}. This long-lived feature was located at 1.8–2.5 bars in 2007 and has been speculated to be linked to a vortex in the same way as the BNC \citep{depater2011}. It is therefore reasonable to deduce that, despite the clouds being located slightly deeper, similar dynamical processes could be at work at this location. The drift speeds of the Berg are similar to those of the BNC as they are located at similar latitudes of opposite hemispheres. It is impossible to deduce which feature could be a cause, or if it is a combination of the two, without spatial resolution at mid-infrared wavelengths.

\section{Conclusion and Future Work}
\label{S:conc}

We have performed a new reduction of Spitzer/IRS observations of Uranus spanning 2004-2007, and adapted the NEMESIS spectral inversion algorithm to provide a full, simultaneous retrieval from 7-21 $\mu$m disc-averaged spectra of Uranus.  This work builds upon that of \citet{orton2014mid1, orton2014mid2}, where an average over multiple longitudes was used to characterise the vertical temperature and composition. The longitudinal variation in emission detected at Uranus during the 2007 equinox can be explained as a physical change in the stratosphere of the planet, possibly due to a localised cold feature associated with a tropospheric vortex and its bright companion cloud, or a stratospheric wave.
\begin{enumerate}
	\item We detect emission variability of up to 15\% at stratospheric altitudes sensed by the hydrocarbon species near the 0.1-mbar pressure level. The tropospheric hydrogen-helium continuum, and the monodeuterated methane that also arises from these deeper levels, both exhibit a negligible variation smaller than 2\%, constraining the phenomenon to the stratosphere.
	\item Spectral inversions show that variations can be explained solely by changes in stratospheric temperatures. A temperature change of less than 3 K is needed to model the observed variation. This is supported by results from high-resolution forward models (primarily sounding the ethane and acetylene emission) constructed using the parameters retrieved from the low-resolution spectra.
	\item Keck II NIRC2 images acquired four days before the Spitzer spectra show a pattern of convective meteorological activity in the upper troposphere in the hemisphere displaying the coolest stratospheric emission, consistent with the possibility of large-scale uplift modulating stratospheric temperatures. One of the bright features observed could be associated with a possible dark spot that is causing an intense and localised cold region like that observed at Neptune. VLT/VISIR 2009 observations also show potential signs of variable emission at mid-latitudes that should motivate future observations.
	\item Spectra from 2005 Spitzer observations show a variability present that is similar, but slightly weaker, than that observed in 2007. Data from 2004 have insufficient signal-to-noise to draw conclusions about the variability at that epoch. 
\end{enumerate}

Future work will include analysis of similar Spitzer observations of Neptune, where preliminary assessments show a lack of significant longitudinal variation. This is inconsistent with the popular view that Neptune is the more meteorologically active of the two planets.

The Spitzer IRS data can provide much detail but without accompanying spatial resolution it is impossible to come to a definitive conclusion as to the origins of the rotational variability. The James Webb Space Telescope, when it launches in 2021, will provide much improved spectral and spatial resolution needed in the mid-infrared band to provide answers to the causes of the observed variation \citep{norwood2016,moses2018}.

\section*{Data Availability}
The processed data used for this investigation is available at \url{https://doi.org/10.5281/zenodo.4617490}. The raw data is accessible from the Spitzer Heritage Archive at \url{https://sha.ipac.caltech.edu/applications/Spitzer/SHA} and can be searched by 'Program', where the 2007 data has program ID 467, the 2005 data has program ID 3534 and the 2004 data has program ID 71.

\section*{Acknowledgments}
Rowe-Gurney, Fletcher and Roman were supported by a European Research Council Consolidator Grant (under the European Union's Horizon 2020 research and innovation programme, grant agreement No 723890) at the University of Leicester.  This work is based, in part, on observations made with the Spitzer Space Telescope, which is operated by the Jet Propulsion Laboratory, California Institute of Technology under a contract with NASA. This research used the ALICE High Performance Computing Facility at the University of Leicester. Orton was supported by funds from NASA distributed to the Jet Propulsion Laboratory, California Institute of Technology. Moses acknowledges support from NASA Solar System Workings grant 80NSSC19K0536. De Pater acknowledges NASA grant NNX16AK14G through the Solar System Observations (SSO) program to the University of California, Berkeley.


\bibliographystyle{elsarticle-harv}

\bibliography{uranus.bib}


\end{document}